\documentclass[aps,prd,twocolumn,amsmath,amssymb,floatfix,nofootinbib,superscriptaddress]{revtex4-1}

\usepackage[utf8]{inputenc}
\usepackage{xcolor}
\usepackage{amsmath}
\usepackage{amssymb}
\usepackage{graphicx}
\usepackage{natbib}
\usepackage{hyperref}
\usepackage{aas_macros}
\usepackage[normalem]{ulem}
\usepackage{siunitx}
\usepackage{comment}
\usepackage[maxfloats=256]{morefloats}
\usepackage{isotope}
\usepackage{bm}
\usepackage{multirow}

\graphicspath{{./figures/}}

\newcommand{\jrcomment}[1]{{\color{green}{[JR: #1]}}}

\newcommand{\lcdm}{$\Lambda$CDM}
\newcommand{\om}{\Omega_{m}}
\newcommand{\Ol}{\Omega_{\Lambda}}
\newcommand{\wx}{w_{\rm X}}

\newcommand{\smu}{Department of Physics,
Southern Methodist University, 3215 Daniel Ave, Dallas, Texas 75205, USA}

\newcommand{\caps}{Center for AstroPhysical Surveys, National Center for Supercomputing Applications, Urbana, IL, 61801, USA}

\begin{document}

\title{Beyond Fisher Forecasting for Cosmology}

\author{Joseph~Ryan}
\email[]{jwryan@smu.edu}
\affiliation{\smu}
\author{Brandon~Stevenson}
\affiliation{\smu}
\author{Cynthia~Trendafilova}
\affiliation{\smu}
\affiliation{\caps}
\author{Joel~Meyers}
\affiliation{\smu}

\date{\today}

\begin{abstract}
The planning and design of future experiments rely heavily on forecasting to assess the potential scientific value provided by a hypothetical set of measurements. The Fisher information matrix, due to its convenient properties and low computational cost, provides an especially useful forecasting tool. However, the Fisher matrix only provides a reasonable approximation to the true likelihood when data are nearly Gaussian distributed and observables have nearly linear dependence on the parameters of interest. Also, Fisher forecasting techniques alone cannot be used to assess their own validity. Thorough sampling of the exact or mock likelihood can definitively determine whether a Fisher forecast is valid, though such sampling is often prohibitively expensive. We propose a simple test, based on the Derivative Approximation for LIkelihoods (DALI) technique, to determine whether the Fisher matrix provides a good approximation to the exact likelihood. We show that the Fisher matrix becomes a poor approximation to the true likelihood in regions where two-dimensional slices of level surfaces of the DALI approximation to the likelihood differ from two-dimensional slices of level surfaces of the Fisher approximation to the likelihood. We demonstrate that our method accurately predicts situations in which the Fisher approximation deviates from the true likelihood for various cosmological models and several data combinations, with only a modest increase in computational cost compared to standard Fisher forecasts.
\end{abstract}

\maketitle

\section{Introduction}
\label{sec:Introduction}

Parameter forecasts play an essential role in the planning, design, and development of future experiments.  For any given model and hypothetical measurement uncertainties, forecasting allows a determination of the precision with which any model parameter can be measured by an experiment before data are collected.  The Fisher information matrix~\cite{1920MNRAS..80..758F,Fisher:1922saa,1925PCPS...22..700F,10.2307/2342435} provides a particularly useful and efficient tool for forecasting. 

The Fisher matrix is straightforward to compute and requires only $\mathcal{O}(N)$ likelihood evaluations for a model with $N$ parameters. The inverse of the Fisher matrix provides a lower bound on the variances and covariances of any unbiased estimators of model parameters, allowing for the analytic determination of marginalized posterior variances and constraint contours.  Due largely to these virtues, Fisher forecasting has become ubiquitous in cosmology~\cite{Jungman:1995bz,Tegmark:1996bz,Bond:1997wr}.

On the other hand, Fisher forecasting suffers from a number of limitations.  The Fisher matrix only provides an exact description when the posterior is a Gaussian function of the model parameters.  This condition is not met unless the data are Gaussian distributed and the observables depend linearly on the model parameters.  Due to the assumption of Gaussianity, the Fisher matrix always predicts symmetric, elliptical constraint contours.  When data are abundant and model parameters are well constrained, the Fisher approximation often provides a good description.  However, significant parameter degeneracy or weakly constraining data can lead to cases where Gaussianity provides a poor approximation; the Fisher matrix will fail to produce accurate forecasts in these circumstances.

Forecasting techniques that do not assume a Gaussian posterior include grid sampling~\cite{Tegmark:2000db}, Markov Chain Monte Carlo (MCMC) methods~\cite{Christensen:2001gj,Lewis:2002ah,Audren:2012wb,Lewis:2013hha}, and nested sampling~\cite{2004AIPC..735..395S,Skilling:2006gxv,Feroz:2007kg,Feroz:2008xx}. These methods provide a more robust determination of non-Gaussian posteriors and parameter constraints, but they are much more computationally expensive than Fisher forecasts, requiring many evaluations of the likelihood function for each model and experimental configuration for which a forecast is desired.

\begin{figure}
    \centering
    \includegraphics[width=\columnwidth]{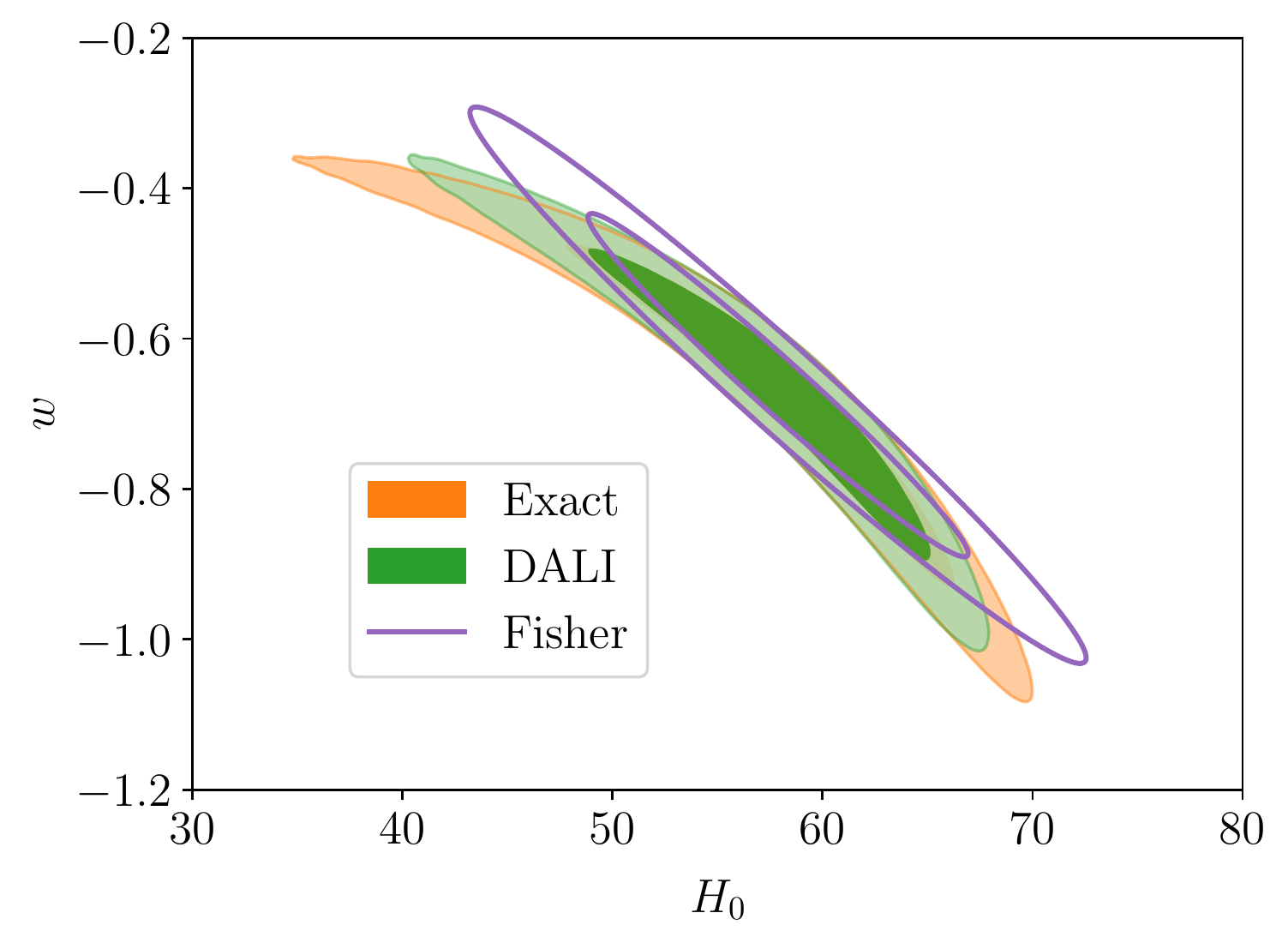}
    \caption{An example of the breakdown of the Fisher approximation. Figure shows constraint contours computed from the exact likelihood (orange), Fisher approximation (purple), and DALI approximation (green) for a simple $w$CDM model fitted to baryon acoustic oscillation (BAO) data. See Sec. \ref{sec:Data_and_Forecasts} for descriptions of $w$CDM and of the BAO data.
    Here and elsewhere, $H_0$ is shown in units of ${\rm km}\hspace{1mm}{\rm s^{-1}}\hspace{1mm}{\rm Mpc^{-1}}$.
    }
    \label{fig:wCDM_example}
\end{figure}

An alternative is provided by the Derivative Approximation for LIkelihoods (DALI) technique~\cite{Sellentin:2014zta,Sellentin:2015axa}, which can be thought of as a non-Gaussian generalization of Fisher matrix methods~\cite{Heavens:2016slh}.  The DALI method is much more efficient than the sampling methods mentioned here, requiring only $\mathcal{O}(N^2)$ likelihood evaluations when forecasting constraints on $N$ parameters.  The DALI method allows for some non-Gaussian aspects of the posterior to be captured, and is capable of producing curved and asymmetric constraint contours (see Figure \ref{fig:wCDM_example} for an example of curved constraint contours).  However, the DALI method generally requires numerical integration to calculate marginalized constraints, in contrast to Fisher forecasts for which marginalization can be achieved analytically.  This numerical integration can be problematic because the forecasted constraints will depend on the details of the numerical integration, including sensitivity to the convergence criterion and the choice of priors on the parameters. Numerical integration also adds computation time; while it is typically a modest computational cost for models with few parameters, it can be a significant burden for models with many parameters.  Furthermore, as we will discuss below, the DALI method is not always guaranteed to improve upon Fisher forecasts. In some cases the DALI method provides estimates which deviate from the true likelihood more than the Fisher approximation. 

A challenge of utilizing Fisher forecasts is that the Fisher matrix does not contain the information necessary to determine whether the Gaussian approximation inherent in the technique is justified.  One therefore needs to employ a technique that goes beyond the Fisher approximation in order to determine its validity.  This could be accomplished with techniques like MCMC forecasts, but this would become prohibitively expensive to carry out for every model and every experimental configuration where a forecast is desired. One of the main goals of this paper is to demonstrate a means by which we can leverage the DALI method to quickly check the validity of Fisher forecasts, without needing to do the numerical integration required to produce a full set of marginalized constraint contours in the DALI method.

Previous studies have demonstrated limitations of Fisher forecasting and identified some sources for its inaccuracies in certain situations~\cite{Joachimi:2011iq,Wolz:2012sr,Khedekar:2012sh,Rodriguez:2013mla,Bellomo:2020pnw,Bernal:2020pwq,Wang:2022kia}.
We rely here on a method of testing the validity of the Fisher technique that should apply quite generally. Our approach utilizes the DALI method to determine when the Fisher approximation fails to accurately match the true likelihood.
Various schemes for utilizing the DALI methods to more precisely approximate weakly non-Gaussian likelihoods have also been developed~\cite{Rizzato:2022hbu,Rover:2022mao}.

This paper is organized as follows.  In Section~\ref{sec:DALI_Formalism} we review the DALI formalism~\cite{Sellentin:2014zta,Sellentin:2015axa}.  In Section~\ref{sec:Curvature} we develop a test to determine whether Fisher forecasts are expected to receive non-trivial corrections, based on the shapes of two-dimensional cross-sections of the posterior estimated with the DALI method.  We then go on to demonstrate the method using cosmological forecasts using expansion history observations, the cosmic microwave background (CMB) power spectrum, and primordial abundance measurements.  We describe the details of the forecasts in Section~\ref{sec:Data_and_Forecasts} and present the results in Section~\ref{sec:Results}.  We comment on some limitations of the DALI method in Section~\ref{sec:Limitations_of_DALI} and conclude in Section~\ref{sec:Conclusion}.

\section{DALI Formalism}
\label{sec:DALI_Formalism}

The starting point for both the Fisher and DALI forecasting methods is a Taylor series expansion of the log-likelihood $\mathcal{L}=-\log{P}$ in the model parameters $p^\alpha$ about the best-fit point defined by $p^{\alpha}_{\rm fid}$, where ``$\log$'' refers to the natural logarithm. Here, $P$ is the posterior, and $\alpha$ is an index that runs from 1 to $N$, where $N$ is the number of parameters. The fiducial best-fit point is typically chosen based on knowledge of the fit of the model to some existing set of observational data.  The Fisher approximation treats this expansion to the lowest non-trivial order, resulting in an approximate log-likelihood which is second-order in the difference between model parameters $p^{\alpha}$ and their fiducial best-fit values $p^{\alpha}_{\rm fid}$ (since first derivatives of the likelihood vanish at the best-fit point).
The posterior in the Fisher approximation is then a Gaussian, which can be written as
\begin{equation}
    P \approx \mathcal{N}_0 {\rm exp}\left[-\frac{1}{2}F_{\alpha \beta}\Delta p^{\alpha}\Delta p^{\beta}\right] \, ,
    \label{eq:Fisher_approx}
\end{equation}
where repeated indices are summed, $\mathcal{N}_0$ is a normalization constant, $\Delta p^{\alpha} := p^{\alpha} - p^{\alpha}_{\rm fid}$, 
and
\begin{equation}
    \label{eq:F_def}
    F_{\alpha \beta} := \boldsymbol{\mu}_{, \alpha}M\boldsymbol{\mu}_{, \beta} \, .
\end{equation}
In Eq.~\eqref{eq:F_def}, $\boldsymbol{\mu}$ is a vector of the observables predicted by a given model evaluated at the fiducial values of the model's parameters, and $M := C^{-1}$ is the inverse of the covariance matrix of the data. We assume for simplicity that the covariance matrix does not depend on the model parameters. The vectors $\boldsymbol{\mu}_{\alpha}$ are contracted with $M$ in the data space, so that the components $F_{\alpha \beta}$ of the Fisher matrix can also be written as $F_{\alpha\beta} = \mu^{i}_{, \alpha}M_{ij}\mu^{j}_{, \beta}$, where Latin indices run from 1 to $n$, $n$ being the number of data points.  There is no difference, in this paper, between tensor components with upper and lower indices.

The Derivative Approximation for LIkelihoods (DALI) method treats the Taylor series expansion of the posterior at higher order, thereby accounting for non-Gaussian parameter dependence in the posterior \cite{Sellentin:2014zta}.  A straightforward expansion in terms of $\Delta p^\alpha$ leads to divergent approximations of the posterior, but the series can be rearranged in order of derivatives of the observables $\boldsymbol{\mu}$ such that the posterior is normalizable and positive-definite at every order.
The ``doublet-DALI" approximation of the posterior is the lowest order correction to the Fisher approximation, and it is given by
\begin{equation}
    \label{eq:doublet_DALI}
    \begin{aligned}
    P = \mathcal{N}_1 {\rm exp}\bigg[& -\frac{1}{2}F_{\alpha \beta}\Delta p^{\alpha}\Delta p^{\beta} - \frac{1}{2}G_{\alpha \beta \gamma}\Delta p^{\alpha} \Delta p^{\beta} \Delta p^{\gamma}\\
    & - \frac{1}{8}H_{\alpha \beta \gamma \delta}\Delta p^{\alpha} \Delta p^{\beta} \Delta p^{\gamma} \Delta p^{\delta}\bigg] + O(3),\\
    \end{aligned}
\end{equation}
where
\begin{equation}
    \label{eq:G_def}
    G_{\alpha \beta \gamma} := \boldsymbol{\mu}_{, \alpha \beta} M \boldsymbol{\mu}_{, \gamma} \, ,
\end{equation}
and
\begin{equation}
    \label{eq:H_def}
    H_{\alpha \beta \gamma \delta} := \boldsymbol{\mu}_{, \alpha\beta}M\boldsymbol{\mu}_{, \gamma\delta} \, ,
\end{equation}
and the notation $O(3)$ refers to terms higher than second order in the derivatives of $\boldsymbol{\mu}$. The DALI expansion can be carried out to any order, but in this paper we focus on the Fisher and the doublet-DALI approximations, since the doublet-DALI approximation suffices for testing the validity of the Fisher approximation, as will be described in the next section.  In the rest of this paper we will refer to the ``doublet-DALI'' approximation as the ``DALI'' approximation.

\section{Fisher validity test}
\label{sec:Curvature}

The Fisher approximation can break down if the observables predicted by a given model depend on the model's parameters in a non-linear fashion. This leads to non-Gaussian correlations among the model parameters represented by terms containing second- or higher-order derivatives of $\boldsymbol{\mu}$. These non-Gaussian correlations manifest themselves in the form of non-ellipsoidal (often banana-shaped) constraint contours when the posterior is plotted in parameter space. Ellipsoidal constraint contours are convex everywhere, while non-ellipsoidal constraint contours can be concave in some regions of the parameter space. We can therefore leverage the DALI method, which is sensitive to non-linear parameter dependence of the posterior, to signal the breakdown of the Fisher approximation.

Qualitatively, this can be achieved by choosing any two of the $N$ parameters $\{p^{\alpha}\}$, and assigning the remaining $N-2$ parameters their fiducial values. This is equivalent to taking a two-dimensional cross-section of the $(N-1)$-dimensional constant-$P$ hypersurface, on which one can plot the level curves of the functions
\begin{equation}
\label{eq:F_level}
    C = \frac{1}{2}F_{\alpha \beta}\Delta p^{\alpha}\Delta p^{\beta}
\end{equation}
and
\begin{equation}
\label{eq:G_level}
\begin{aligned}
    C = \frac{1}{2}F_{\alpha \beta}\Delta p^{\alpha} \Delta p^{\beta} & + \frac{1}{2}G_{\alpha \beta \gamma}\Delta p^{\alpha} \Delta p^{\beta} \Delta p^{\gamma}\\
    & + \frac{1}{8} H_{\alpha \beta \gamma \delta}\Delta p^{\alpha} \Delta p^{\beta} \Delta p^{\gamma} \Delta p^{\delta}\\
\end{aligned}
\end{equation}
for various values of $C$. Because Eq.~\eqref{eq:G_level} describes the level curves of the DALI approximation, and Eq.~\eqref{eq:F_level} describes the level curves of the Fisher approximation, one just needs to visually examine the extent to which the level curves of Eq.~\eqref{eq:G_level} deviate from those of Eq.~\eqref{eq:F_level} to assess the validity of the Fisher approximation. In Sec.~\ref{sec:Results} we show the results of this test.

As an optional supplement to the cross-section test, a quantitative measure that can be used to indicate a difference between the Fisher and DALI approximations to the likelihood hypersurfaces is the extrinsic curvature of the level curves. To compute the extrinsic curvature as a function of model parameters, we define the constraint equation
\begin{equation}
\label{eq:Phi=0_new}
    \Phi\left(p^{\alpha}\right) := \mathcal{L}\left(p^{\alpha}\right) - C = 0 \, ,
\end{equation}
and a unit normal vector
\begin{equation}
\label{eq:n_def_new}
    n_{\alpha} := \frac{\Phi_{, \alpha}}{\sqrt{g^{\alpha \beta}\Phi_{, \alpha}\Phi_{, \beta}}} \, ,
\end{equation}
such that the extrinsic curvature, in two dimensions, is
\begin{equation}
\label{eq:K_2D_new}
    K = n^{\alpha}_{\phantom{\alpha},\alpha} = \frac{\Phi_{,00}\left(\Phi_{,1}\right)^2 + \Phi_{,11}\left(\Phi_{,0}\right)^2 - 2\Phi_{,0}\Phi_{,1}\Phi_{,01}}{\left[\left(\Phi_{,0}\right)^2 + \left(\Phi_{,1}\right)^2\right]^{3/2}}.
\end{equation}
If $K < 0$ at any points on a two-dimensional cross-section of the parameter space, then the two-dimensional likelihood level curves will be concave at those points.   Since the Fisher-approximated level curves are convex everywhere, concavity of the DALI-approximated level curves signals a region in which the two approximations differ (though they can differ for other reasons).  Identifying regions in which the DALI-approximated level curves exhibit concavity can aid in finding cases in which the Fisher approximation breaks down, especially since regions of concavity can be detected without the necessity of making plots. We show the level curves and points where the DALI level curves are concave in Figures~\ref{fig:Fisher_Not_OK_1}-\ref{fig:BBN_30}. Different approaches to the geometry of likelihood contours and its applications have been discussed elsewhere~\cite{Transtrum:2010zz,Transtrum:2011abc,Giesel:2020mwj,Quinn:2021uyo}.

\section{Data and Forecasts}
\label{sec:Data_and_Forecasts}

The sets of observables that we study fall into three categories: measurements of the expansion history in the range $z \leq 2.73$, CMB observations, and measurements of primordial light element abundances. To obtain forecast constraints, we consider predictions of the expansion history data from three models (2-parameter flat \lcdm, 3-parameter non-flat \lcdm, and 3-parameter flat $w$CDM). We study an 8-parameter flat \lcdm\ model with the CMB data, and we examine 3-parameter models of Big Bang Nucleosynthesis (BBN) with the primordial abundance observations.

The reader should keep in mind that our aim is not to use the DALI formalism to obtain new constraints or forecasts on cosmological model parameters. Our models and data should be seen merely as a proving ground for our proposed Fisher validity test. We only intend to show some examples of what can be done with the formalism, in addition to analyzing how and why the DALI approximation breaks down.

\subsection{Expansion history data}
\label{subsec:low_redshift}
The expansion history measurements we use consist of a set of cosmic chronometer and standard ruler data. The cosmic chronometer ($H(z)$) data, collected from \cite{Simon:2004tf, Moresco:2012jh, Stern:2009ep, Moresco:2015cya, Zhang:2012mp, Moresco:2016mzx, Ratsimbazafy:2017vga}, comprise a set of uncorrelated measurements of the Hubble parameter as a function of redshift, and are compiled in \cite{Ryan:2018aif}. 

The standard ruler data can be broken down into two subsets: baryon acoustic oscillation (BAO) and quasar angular size (QSO). The BAO subset consists of eleven distance and Hubble parameter measurements, collected from Refs. \cite{BOSS:2016wmc, Carter:2018vce, DES:2017rfo, Ata:2017dya, duMasdesBourboux:2020pck} and compiled in Table 1 of Ref.~\cite{Cao:2021ldv}. Unlike the $H(z)$ data, subsets of these BAO data are correlated; we use the covariance matrices given by Eq.~(20) in Ref.~\cite{Ryan:2019uor} and Eq.~(8) of Ref.~\cite{Cao:2021ldv} to account for the correlations (see references for further details).

The QSO subset comprises measurements of the angular sizes of 120 intermediate-luminosity quasars, from knowledge of the intrinsic linear size $l_m$ of the quasars within the sample, and measurements of their angular diameter distances at a given redshift. These measurements were taken from Ref.~\cite{Cao:2017ivt}. In contrast to the recent analysis of Ref.~\cite{Cao:2021cix}, we do not vary $l_m$ as a free parameter. We choose instead to follow the practice of Refs.~\cite{Ryan:2018aif, Ryan:2019uor} in setting $l_m = 11.03$~pc while ignoring the relatively small uncertainty of $0.25$~pc. 

We fit three simple models to the data using the Fisher and DALI approximations, and compare these fits to those obtained from the exact (i.e. not approximated using the DALI method) likelihoods. The first, and simplest, model is the 2-parameter \lcdm\ model, with an expansion history characterized by the Hubble parameter
\begin{equation}
    \label{eq:LCDM_H(z)}
    H(z) = H_0\sqrt{\Omega_{m}\left(1 + z\right)^3 + 1 - \om}.
\end{equation}
This model has two parameters: the Hubble constant, $H_0$, and the non-relativistic mass density in units of the critical mass density, $\Omega_{m}$. Throughout this article, ``$\Lambda$CDM'' stands for ``$\Lambda$ Cold Dark Matter'', where $\Lambda$ refers to the cosmological constant, and cold dark matter is the dominant contributor to the nonrelativistic mass density. We also consider the non-flat \lcdm\ model, with the Hubble parameter
\begin{equation}
    H(z) = H_0\sqrt{\om\left(1 + z\right)^3 + \Omega_{k}\left(1 + z\right)^2 + \Ol}.
\end{equation}
This model adds one free parameter compared to 2-parameter \lcdm: the mass density of the cosmological constant $\Lambda$ in units of the critical mass density, $\Ol$. In contrast to 2-parameter \lcdm, the 3-parameter \lcdm\ model can have curved spatial hypersurfaces, with the amount of curvature quantified by the curvature energy density parameter in units of the critical mass density, $\Omega_{k} := 1 - \Omega_{m} - \Omega_{\Lambda}$. Finally, we consider a flat $w$CDM model in which the dark energy is represented by a perfect fluid with equation of state $\frac{p_{\rm X}}{\rho_{\rm X}} = w_{\rm X}$, and a Hubble parameter given by
\begin{equation}
    H(z) = H_0 \sqrt{\om\left(1 + z\right)^3 + \left(1 - \om\right)\left(1 + z\right)^{3\left(1 + \wx\right)}}.
\end{equation}
This model is a simple generalization of flat \lcdm, which allows for simple time variation of the dark energy density through the factor $\left(1 + z\right)^{3\left(1 + \wx\right)}$.\footnote{Because the model has flat spatial hypersurfaces, the dark energy density in units of the critical mass density must obey the constraint $\Omega_{\rm X} = 1 - \Omega_{m}$.} Here the free parameters are $H_0$, $\Omega_{m}$, and $w_{\rm X}$.

To estimate the exact likelihood corresponding to the expansion history data, we use \texttt{Cobaya}~\cite{Torrado:2020dgo,2019ascl.soft10019T}\footnote{\url{https://github.com/CobayaSampler/cobaya}} to sample over the exact likelihoods and the DALI-approximated likelihoods, using \texttt{Cobaya}'s MCMC sampler (adapted from \texttt{CosmoMC}~\cite{Lewis:2002ah,Lewis:2013hha}). For the expansion history data we used a convergence criterion of either $R - 1 = 10^{-5}$ or $R - 1 = 5 \times 10^{-6}$, depending on the model and data combination.

\subsection{CMB forecasts}
\label{subsec:CMB_forecasts}

The unlensed CMB, lensing deflection, and lensed CMB power spectra used in our forecasts are computed with the \texttt{CAMB} Boltzmann code~\cite{Lewis:1999bs,Howlett:2012mh}\footnote{\url{https://camb.info/}}. Lensing reconstruction noise is calculated with \texttt{CLASS\_delens}\footnote{\url{https://github.com/selimhotinli/class_delens/}} according to the implementation detailed in \cite{Hotinli:2021umk}, and Fisher and DALI forecasts are calculated using \texttt{FisherLens}\footnote{\url{https://github.com/ctrendafilova/FisherLens}}.

The two CMB experimental noise configurations that we consider are given in Table~\ref{table:experiments}, and we assume Gaussian noise with temperature power spectrum of the form
\begin{equation}
    N_{\ell}^{TT} = \Delta_T^2 \, \mathrm{exp} \left( \ell(\ell+1) \frac{\theta_{\mathrm{FWHM}}^2}{8\log{2}} \right) \, .
    \label{eq:TT_noise}
\end{equation}
Here $\Delta_T$ is the instrumental noise in $\mu$K-rad, $\theta_{\mathrm{FWHM}}$ is the full-width at half-maximum beam size in radians, and we assume fully polarized detectors such that $N_{\ell}^{EE} = N_{\ell}^{BB} = 2N_{\ell}^{TT}$.

The elements of the Fisher matrix are given by 
\begin{equation}
    F_{\alpha\beta}^{\mathrm{CMB}} = \sum\limits_{\ell_1, \ell_2} \ \sum\limits_{X Y, W Z} 
    \frac{\partial C_{\ell_1}^{XY}}{\partial p^\alpha} 
    \left[ \mathrm{Cov}_{\ell_1\ell_2}^{XY,WZ} \right]^{-1}
    \frac{\partial C_{\ell_2}^{WZ}}{\partial p^\beta} \, ,
    \label{eq:FisherMatrixCMB}
\end{equation}
where $p^\alpha$ are the cosmological parameters of interest, and $XY$, $WZ$ run over $TT$, $TE$, $EE$, and $dd$. The lensing deflection spectrum is $C_\ell^{dd} = \ell(\ell+1)C_\ell^{\phi\phi}$.
The elements of the DALI tensors needed for the doublet-DALI approximation are given by
\begin{equation}
    G_{\alpha\beta\gamma}^{\mathrm{CMB}} = \sum\limits_{\ell_1, \ell_2} \ \sum\limits_{X Y, W Z} 
    \frac{\partial C_{\ell_1}^{XY}}{\partial p^\alpha \partial p^\beta} 
    \left[ \mathrm{Cov}_{\ell_1\ell_2}^{XY,WZ} \right]^{-1}
    \frac{\partial C_{\ell_2}^{WZ}}{\partial p^\gamma} \, 
    \label{eq:DALI3CMB}
\end{equation}
and
\begin{equation}
    H_{\alpha\beta\gamma\delta}^{\mathrm{CMB}} = \sum\limits_{\ell_1, \ell_2} \ \sum\limits_{X Y, W Z} 
    \frac{\partial C_{\ell_1}^{XY}}{\partial p^\alpha \partial p^\beta} 
    \left[ \mathrm{Cov}_{\ell_1\ell_2}^{XY,WZ} \right]^{-1}
    \frac{\partial C_{\ell_2}^{WZ}}{\partial p^\gamma \partial p^\delta} \, .
    \label{eq:DALI4CMB}
\end{equation}

For $TT$ spectra, the sums over $\ell$ go from $\ell_\mathrm{min} = 30$ to $\ell_\mathrm{max} = 3000$, and when performing lensing reconstruction we take $\ell_\mathrm{max} = 3000$ for temperature; for polarization spectra, we take $\ell_\mathrm{max} = 5000$. For the lensing spectrum, we include in the sums the modes from $\ell_\mathrm{min} = 2$ to $\ell_\mathrm{max} = 5000$. We assume a sky fraction of $f_\mathrm{sky} = 0.6$, and we include a prior on $\tau$ with $\sigma_\tau = 0.007$. The cosmological parameters, fiducial values, and step sizes for numerical derivatives used in our forecasts are given in Table~\ref{table:cosmo_fiducial}. We specify the primordial Helium abundance such that it is consistent with predictions from BBN.


\begin{table}[t!]
\renewcommand{\arraystretch}{1.2}
    \begin{center}
     \begin{tabular}{l @{\hskip 12pt} c@{\hskip 12pt}c@{\hskip 12pt}c} 
     \toprule
       Label & $\Delta_T$ ($\mu$K-arcmin)    &   $\theta_{\rm FWHM}$ (arcmin) \\ [0.5ex] 
     \hline
    Experiment A & 5 & 1.4  \\ 
    Experiment B & 1 & 1.4  \\
      \hline
    \end{tabular}
    \caption{
    Noise levels for CMB experiments used in forecasts. Experiment A corresponds approximately to the beam size and white noise level expected from Simons Observatory~\cite{SimonsObservatory:2018koc}, and Experiment B corresponds to that of CMB-S4~\cite{CMB-S4:2016ple,Abazajian:2019eic}.
    }
    \label{table:experiments}
    \end{center}
\end{table}


\begin{table}[t!]
\renewcommand{\arraystretch}{1.2}
\begin{center}
 \begin{tabular}{l@{\hskip 12pt}c@{\hskip 12pt}c} 
 \toprule
   Parameter     &   Fiducial Value      & Step Size     \\ [0.5ex] 
 \hline
   $\Omega_c h^2$ &   0.1197 	            & 0.0030 	    \\ 
   $\Omega_b h^2$ &   0.0222 	            & $8.0\times10^{-4}$ 	    \\
   $\theta_s$     &   0.010409 	            & $5.0\times10^{-5}$ 	    \\
   $\tau$         &   0.060 	            & 0.020 	    \\
   $A_s$          &   $2.196\times10^{-9}$  & $0.1\times10^{-9}$ 	    \\
   $n_s$          &   0.9655 	            & 0.010 	    \\
 $N_\mathrm{eff}$ & 3.046 & 0.080 \\
 $m_{\nu}$ (meV) & 60 & 20 \\
  \hline
\end{tabular}
    \caption{
    Fiducial cosmological parameters and step sizes for numerical derivatives used in forecasts, reproduced from \cite{Allison:2015qca}.
    }
\label{table:cosmo_fiducial}
\end{center}
\end{table}

Gravitational lensing of the primary CMB changes the statistics of the temperature and polarization anisotropies, coupling modes of different $\ell$ and leading to non-Gaussian off-diagonal contributions to the power spectrum covariance matrix. Neglecting these contributions when performing parameter forecasts will lead to overly optimistic estimates of parameter constraints, since one is effectively reducing the number of independent $\ell$ modes~\cite{Hotinli:2021umk}.
In this work, we do not include the lensing-induced non-Gaussian covariances in our forecasts, in order to facilitate consistent comparison with the results of our mock likelihood analysis, 
which assumes Gaussian statistics for the CMB anisotropies. Neglecting the lensing-induced non-Gaussian covariances leads to slightly tighter constraint contours when forecasting using lensed CMB power spectra, but we do not expect it to significantly impact their shape.

\begin{table}[t!]
\renewcommand{\arraystretch}{1.2}
\begin{center}
 \begin{tabular}{c@{\hskip 12pt} c@{\hskip 12pt} | c@{\hskip 12pt} c@{\hskip 12pt} c} 
 \toprule
   Redshift & $\frac{\sigma(r_s/d_V)}{(r_s/d_V)}$ (\%) &&
   Redshift & $\frac{\sigma(r_s/d_V)}{(r_s/d_V)}$ (\%) \\ [0.5ex] 
 \hline
0.15 &   1.89   && 1.05 &   0.59\\ 
0.25 &   1.26   && 1.15 &   0.60\\
0.35 &   0.98   && 1.25 &   0.57\\
0.45 &   0.80   && 1.35 &   0.66\\
0.55 &   0.68   && 1.45 &   0.75\\
0.65 &   0.60   && 1.55 &   0.95\\
0.75 &   0.52   && 1.65 &   1.48\\
0.85 &   0.51   && 1.75 &   2.28\\
0.95 &   0.56   && 1.85 &   3.03\\
  \hline
\end{tabular}
    \caption{
    Expected fractional uncertainties on $r_s/d_V$ from forecast DESI data, reproduced from \cite{Allison:2015qca}. 
    }
\label{table:desi_bao_errors}
\end{center}
\end{table}

In addition to information from CMB forecasts, we also include information from Baryon Acoustic Oscillation (BAO) experiments. The elements of the Fisher information matrix are given by
\begin{equation}
    F_{\alpha\beta}^{\mathrm{BAO}} = \sum\limits_{a}  
    \frac{1}{\sigma_{f,a}^2}
    \frac{\partial f_a}{\partial p^\alpha} 
    \frac{\partial f_a}{\partial p^\beta} \, ,
    \label{eq:FisherMatrixBAO}
\end{equation}
where $f_a \equiv f(z_a) = r_s/d_V(z_a)$; $r_s$ is the sound horizon at photon-baryon decoupling, and $d_V$ is the volume distance to source galaxies at redshifts $z_a$. Similarly to the CMB case, the relevant BAO DALI tensors are given by
\begin{equation}
    G_{\alpha\beta\gamma}^{\mathrm{BAO}} = \sum\limits_{a}  
    \frac{1}{\sigma_{f,a}^2}
    \frac{\partial f_a}{\partial p^\alpha \partial p^\beta} 
    \frac{\partial f_a}{\partial p^\gamma} \, 
    \label{eq:DALI3BAO}
\end{equation}
and
\begin{equation}
    H_{\alpha\beta\gamma\delta}^{\mathrm{BAO}} = \sum\limits_{a}  
    \frac{1}{\sigma_{f,a}^2}
    \frac{\partial f_a}{\partial p^\alpha \partial p^\beta}
    \frac{\partial f_a}{\partial p^\gamma \partial p^\delta} \, .
    \label{eq:DALI4BAO}
\end{equation}
The forecasted fractional errors in $f_a$ for the DESI experiment are given in Table~\ref{table:desi_bao_errors}.

To estimate the exact likelihood corresponding to our CMB experiments defined above, we use \texttt{Cobaya}
to sample over the posterior.
Lensed CMB power spectra and lensing spectra are computed using the \texttt{CAMB} cosmological Boltzmann code \cite{Lewis:1999bs,Howlett:2012mh}. Our likelihood code is based on the mock CMB likelihood from \texttt{MontePython}~\cite{Audren:2012wb}.

We define the log-likelihood function as \cite{Hamimeche:2008ai}
\begin{align}\label{eq:CMBlike}
    \mathcal{L} (\{\bm{C}_\ell\}|\{\hat{\bm{C}_\ell}\}) = &
    -\frac{1}{2} \sum_{\ell} (2\ell+1) f_\mathrm{sky} \nonumber \\
    &\times \left\{ \mathrm{Tr} [\hat{\bm{C}_\ell} \bm{C}_\ell^{-1}] - 
    \ln |\hat{\bm{C}_\ell} \bm{C}_\ell^{-1}| - k \right\},
\end{align}
where $k$ counts the number of correlated Gaussian fields included in the forecast. 
$\hat{\bm{C}_\ell}$ corresponds to simulated CMB power spectra calculated according to our fiducial values and representing our mock experimental data. $\bm{C}_\ell$ are functions of the cosmological parameters, sampled over by \texttt{Cobaya}. When using temperature, $E$-mode polarization, and lensing information, we have $k=3$, and the $\bm{C}_\ell$ become 3$\times$3 matrices given by

\begin{equation}\label{eq:Cmatrix}
    \bm{C}_\ell = \begin{bmatrix}
            C_\ell^{TT} + N_\ell^{TT} & C_\ell^{TE} & 0\\
            C_\ell^{TE} & C_\ell^{EE} + N_\ell^{EE} & 0\\
            0 & 0 & C_\ell^{dd} + N_\ell^{dd} \end{bmatrix}.
\end{equation}
The $\hat{\bm{C}_\ell}$ matrices have the same functional form in terms of their respective CMB anisotropy and noise power spectra. The sky fraction and fiducial parameter values are the same as those used in the Fisher and DALI forecasts, and we include a Gaussian prior on $\tau$ with $\sigma_\tau = 0.007$.

We use the same $\ell$ ranges in all forecasts, with the $\bm{C}_\ell$ and $\hat{\bm{C}_\ell}$ matrices taking a different form depending on the value of $\ell$ in the sum in Eq.~\eqref{eq:CMBlike}. For $2 \leq \ell < 30$,
\begin{equation}
    \bm{C}_\ell = \begin{bmatrix}
            C_\ell^{dd} + N_\ell^{dd} \end{bmatrix}.
\end{equation}
For $30 \leq \ell \leq 3000$, the matrices take the form given in Eq.~\eqref{eq:Cmatrix}. 
Finally, for $3000 < \ell \leq 5000$, we want to exclude $TT$ information, while keeping $TE$ and $EE$. We accomplish this by setting the $TT$ entry to be the same in both $\bm{C}_\ell$ and $\hat{\bm{C}_\ell}$, in both cases using the simulated CMB power spectrum representing mock data, so that the difference does not contribute to the likelihood.

For BAO data, our log-likelihood function is defined as
\begin{align}
    \mathcal{L} & \left(\left\{\frac{r_s}{d_V}\right\}\left|\left\{\frac{\hat{r}_s}{d_V}\right\}\right)\right.  \nonumber \\
    &= -\frac{1}{2}\sum_z \left(\frac{r_s}{d_V} - \frac{\hat{r}_s}{d_V} \right)^2  \left(\sigma\left(\frac{r_s}{d_V}\right) \right)^{-2}.
    \label{eq:BAO_LogLike}
\end{align}
From \texttt{CAMB}, we get the angular diameter distance, $d_A$, the Hubble rate, $H$, and the radius of the sound horizon at the drag epoch, $r_s(z_d)$, and we calculate $r_s/d_V$ with
\begin{equation}
    {d_V}(z) = \left[ c z(1+z)^2 d_A^2(z) H^{-1}(z) \right]^{1/3}
\end{equation}
at each of the redshifts specified in Table~\ref{table:desi_bao_errors}~\cite{Allison:2015qca}. Whether our mock likelihood forecasts include BAO or only CMB data, we use a convergence criterion of $R-1 = 0.01$.

 \subsection{Primordial abundance measurements}
\label{subsec:BBN_forecasts}

We utilize observations of the light element abundances to place constraints on the baryon-to-photon ratio, $\eta_{10} \equiv \eta \times 10^{10}$, the light relic density, $N_\nu$, and the neutron lifetime, $\tau_n$. We focus on observations of the primordial abundance of deuterium, $\isotope{D}/\isotope{H}$, and the primordial mass fraction of helium-4, $Y_p$. 

To predict the primordial abundances for a given set of cosmological parameters, we use the publicly available \texttt{PArthENoPE} software\footnote{\url{http://parthenope.na.infn.it/}}~\cite{Gariazzo:2021iiu, Consiglio:2017pot, Pisanti:2007hk}. We utilize the standard settings for \texttt{PArthENoPE} discussed in \cite{Pisanti:2007hk} with a 9-nuclide nuclear network, with nuclear rates fixed at their best-fit values.  Our choice to fix nuclear rates will give errors that are slightly tighter than other published results. We believe this choice is justified because, as mentioned in Sec.~\ref{sec:Data_and_Forecasts}, our goal is not to present fully robust constraints, but to test the validity of the Fisher approximation.

We found that numerical second derivatives of the primordial abundances predicted by \texttt{PArthENoPE} were not sufficiently stable for use in our analysis.
In order to calculate the derivatives used in the Fisher matrix and DALI tensors, we therefore used the semi-analytic parametrization of the light element abundances from Ref.~\cite{Yeh:2020mgl}
\begin{equation}
    Y_p = 0.24695 \left( \frac{\eta_{10}}{6.123}\right)^{0.039}\left( \frac{N_\nu}{3.0}\right)^{0.163}\left( \frac{\tau_n}{879.4~\mathrm{s}}\right)^{0.729} \, ,
    \label{eq:Yp_analytic}
\end{equation}
and
\begin{align}
    \frac{\isotope{D}}{\isotope[]{H}} & =  2.493\times 10^{-5}   \nonumber \\
     &  \times \left( \frac{\eta_{10}}{6.123}\right)^{-1.634}  \left( \frac{N_\nu}{3.0}\right)^{0.405}\left( \frac{\tau_n}{879.4~\mathrm{s}}\right)^{0.418} \, .
    \label{eq:DH_analytic}
\end{align}

The observational errors on the primordial abundances are assumed to be Gaussian and are taken to be $\sigma(Y_p) = 0.0040$ from Ref.~\cite{Aver:2015iza} and  $\sigma(\isotope{D}/\isotope{H}) = 0.030 \times 10^{-5}$ from Ref.~\cite{Cooke:2017cwo}.  We consider two different uncertainties for the neutron lifetime discussed below.  The exact likelihood follows a form analogous to Eq.~\eqref{eq:BAO_LogLike}.
We used \texttt{Cobaya}  to sample over model parameters for both the DALI approximation and the exact likelihood, using \texttt{PArthENoPE} to compute the primordial abundances at each sampled point for the exact likelihood.
A convergence criterion of $R-1=0.005$ was used for both the DALI and exact likelihood sampling.

\section{Cross-section Test Results}
\label{sec:Results}

Here we apply the cross-section test described in Sec.~\ref{sec:Curvature} to the models and data sets described in Sec.~\ref{sec:Data_and_Forecasts}.

\subsection{Expansion history data}
\label{subsec:low-z_results}

\begin{figure}[tbhp]
    \centering
    \includegraphics[width=\columnwidth]{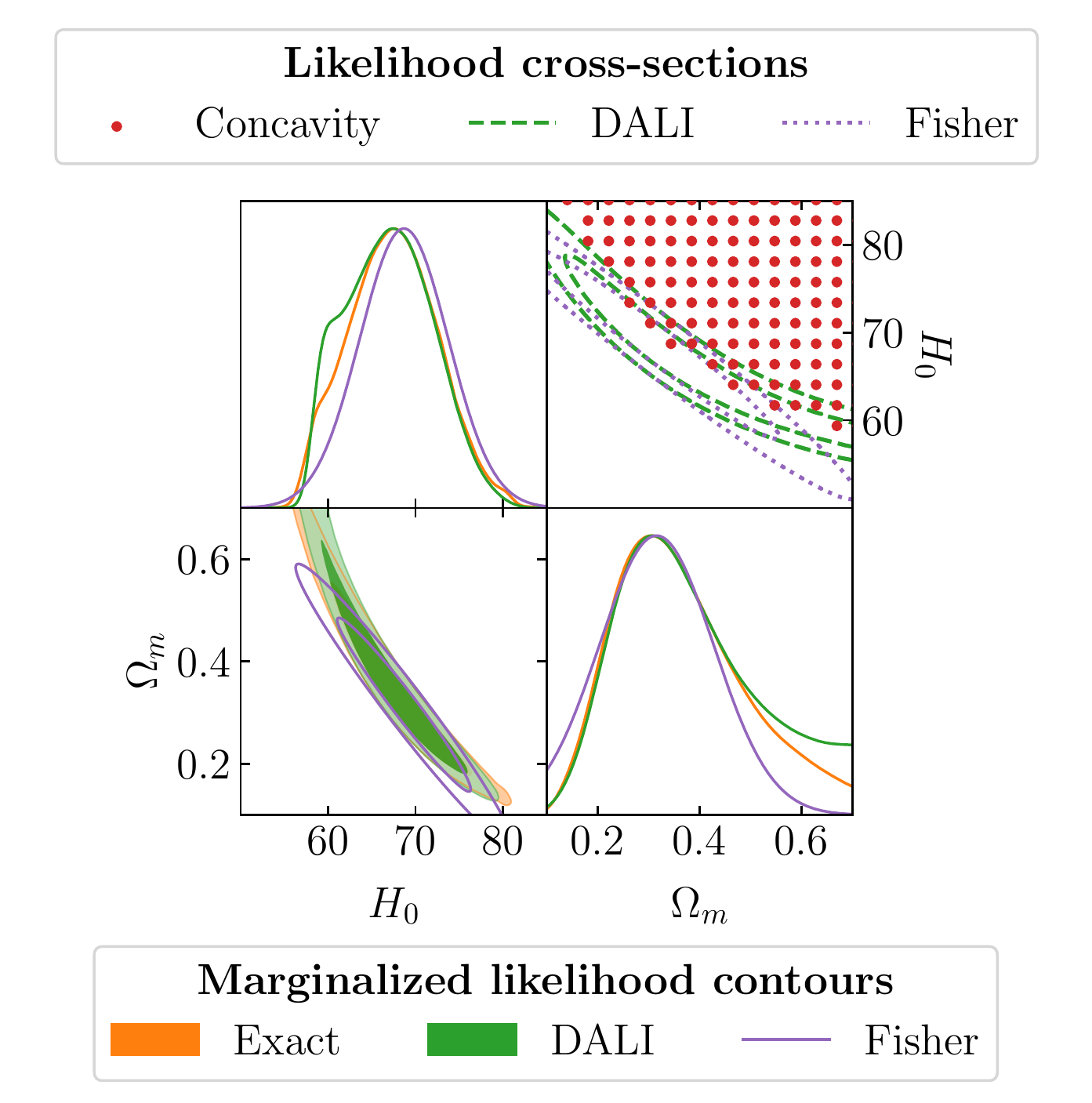}
    \caption{The lower triangle shows the fit of the exact likelihood, DALI, and Fisher approximations to QSO data in the flat $\Lambda$CDM model. Orange and green filled contours in the lower left panel correspond to 1- and 2-$\sigma$ confidence contours for the exact likelihood and DALI approximation, respectively, and purple unfilled contours correspond to 1- and 2-$\sigma$ confidence contours for the Fisher approximation. The upper right panel shows a two-dimensional slice through the 1- and 2-$\sigma$ likelihood hypersurfaces at the fiducial best-fit point. The purple dotted curves represent the slice through the Fisher hypersurface, dashed green curves represent the slice through the DALI hypersurface, and red dots indicate points at which the slice through the DALI hypersurface is concave (which occurs wherever $K < 0$).}
    \label{fig:Fisher_Not_OK_1}
\end{figure}

\begin{figure}[tbhp]
    \centering
    \includegraphics[width=\columnwidth]{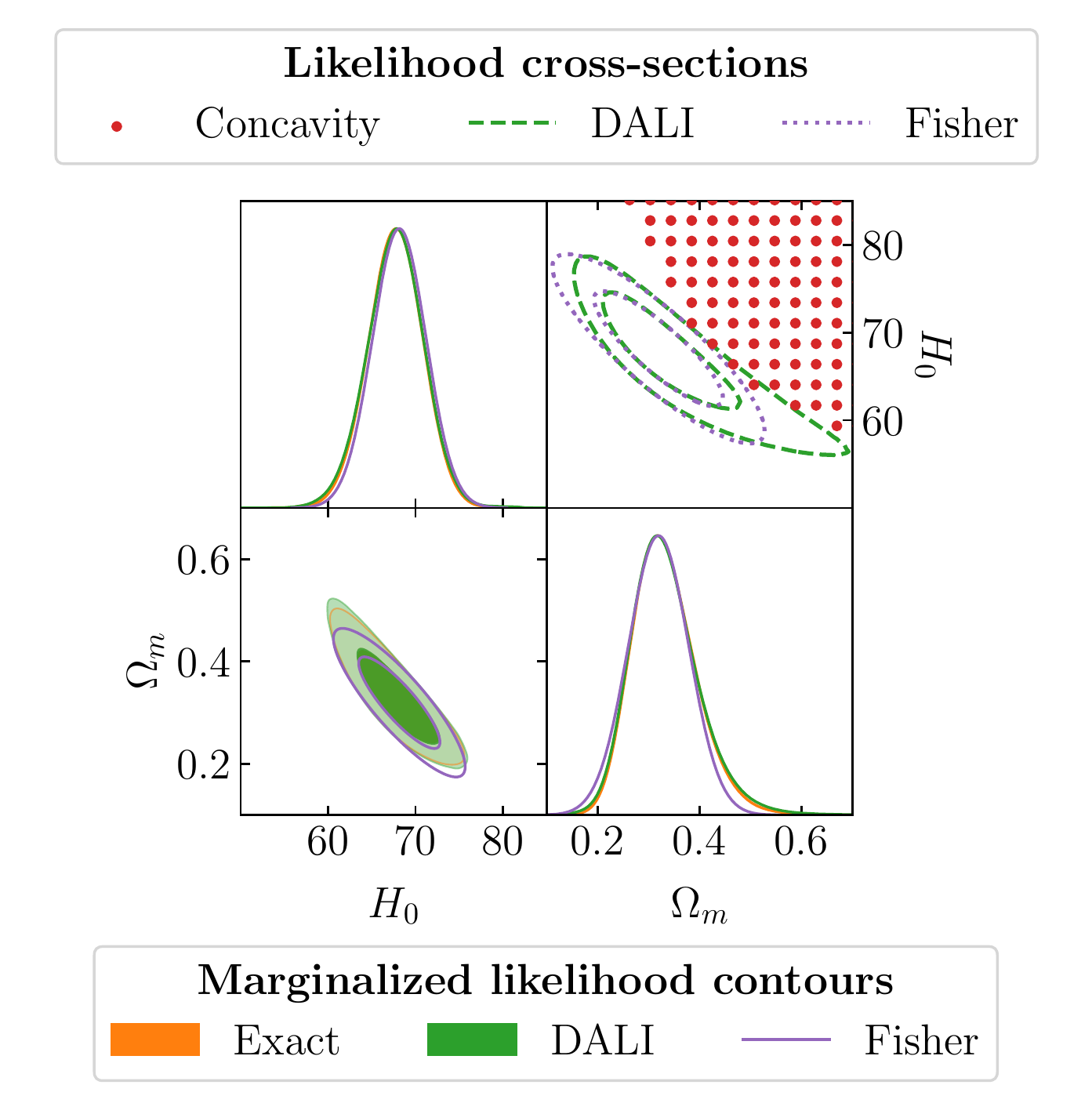}
    \caption{The same as Figure \ref{fig:Fisher_Not_OK_1}, but using $H(z)$ data to fit the flat $\Lambda$CDM model. 
    }
    \label{fig:Fisher_Not_OK_2}
\end{figure}

\begin{figure}[tbhp]
    \centering
    \includegraphics[width=\columnwidth]{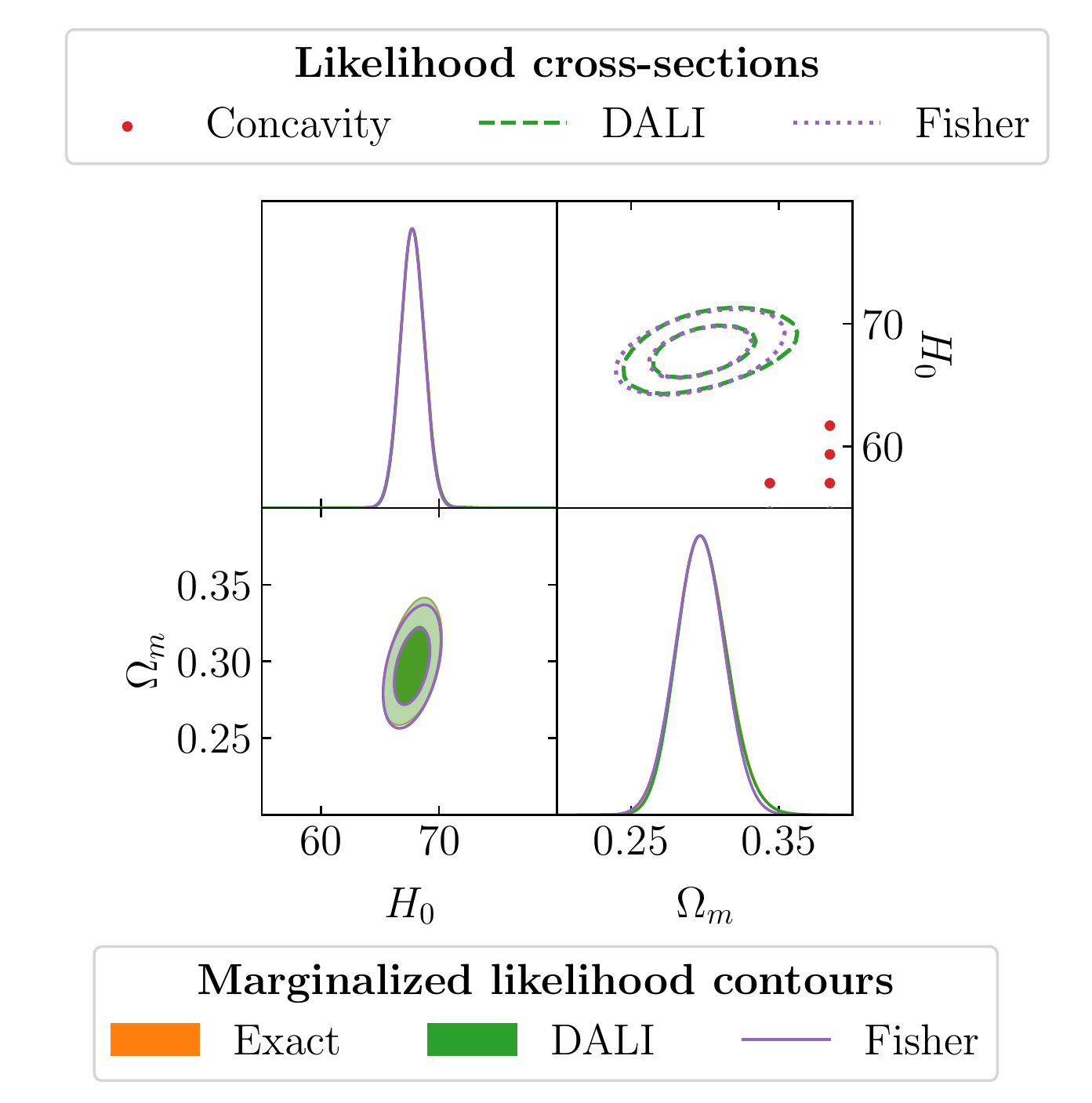}
    \caption{The same as Figure \ref{fig:Fisher_Not_OK_1}, but using BAO data to fit the flat $\Lambda$CDM model. 
    }
    \label{fig:Fisher_OK}
\end{figure}

\begin{figure}[tbhp]
    \centering
    \includegraphics[width=\columnwidth]{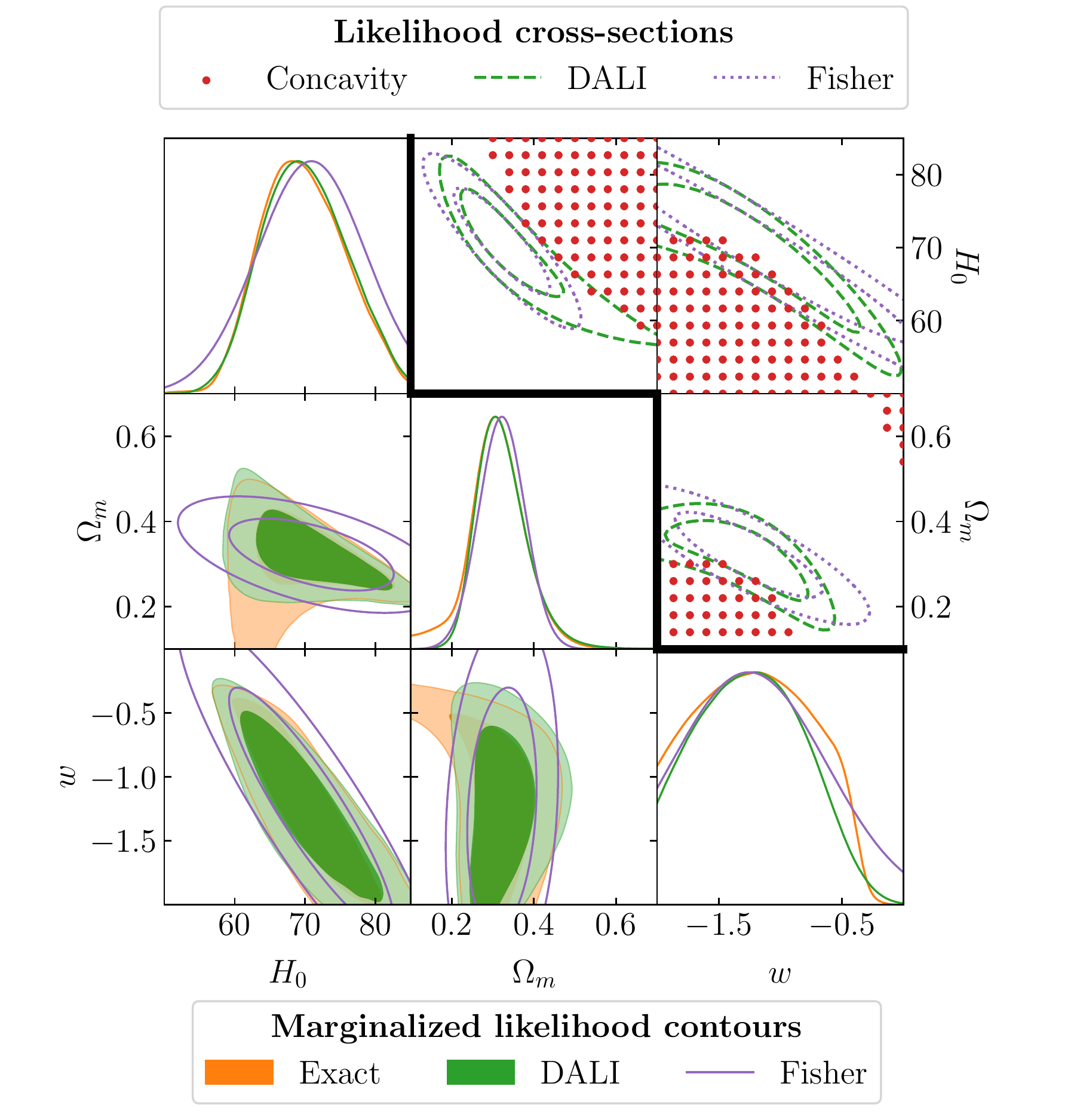}
    \caption{The lower triangle, below the bold ``staircase'' running through the middle of the figure, shows the one- and two-dimensional marginalized constraints on the parameters of the flat $w$CDM model fitted to $H(z)$ data. The upper triangle shows two-dimensional cross-sections through the Fisher- and DALI-approximated likelihoods, where the red dots represent points at which the DALI-approximated likelihood is concave. All other colors are the same as in Figures \ref{fig:Fisher_Not_OK_1} - \ref{fig:Fisher_OK}. 
    }
    \label{fig:FXCDM_Hz}
\end{figure}

\begin{figure}[t!]
    \centering
    \includegraphics[width=\columnwidth]{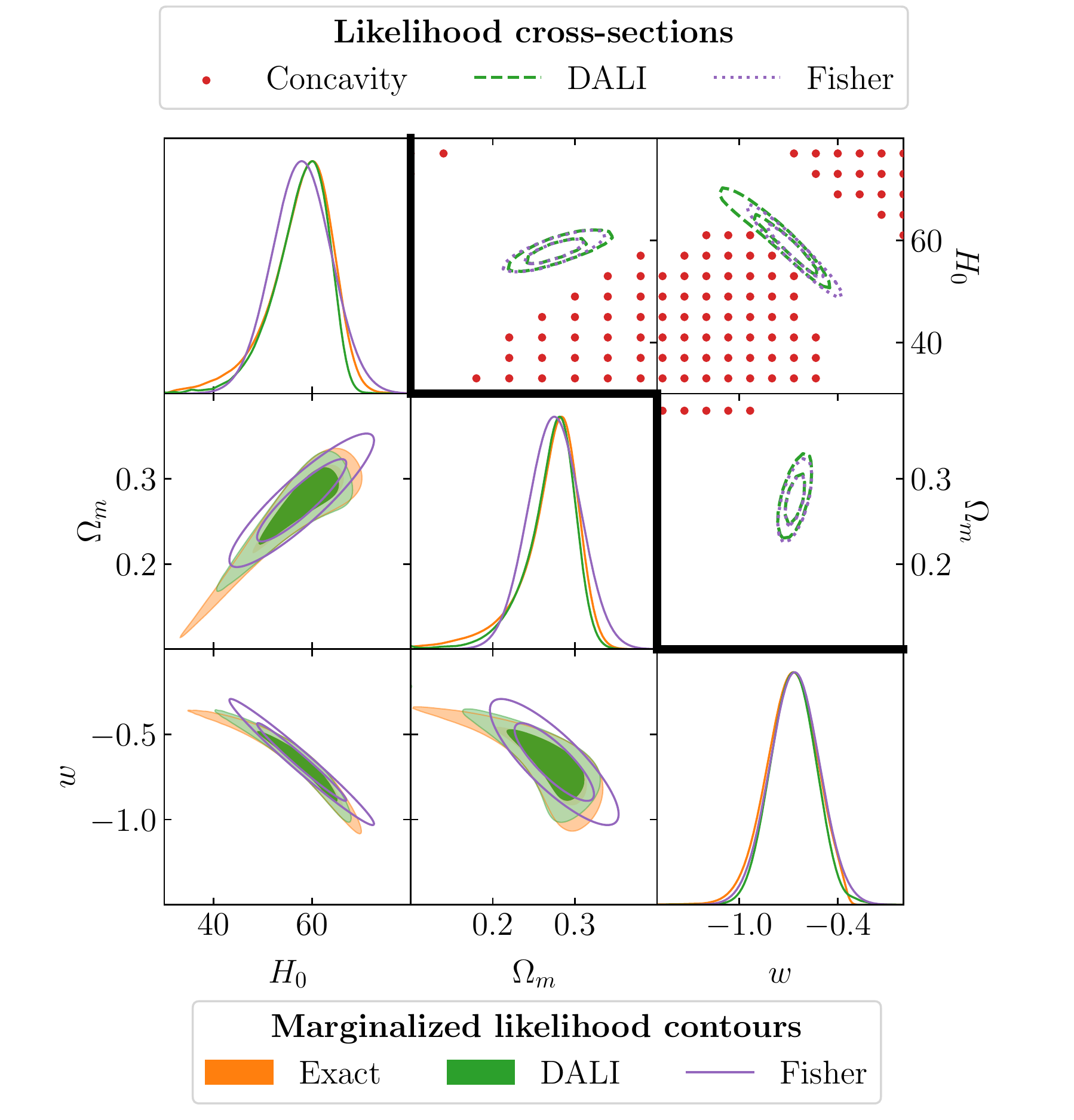}
    \caption{The same as Figure \ref{fig:FXCDM_Hz}, but using BAO data to fit the flat $w$CDM model. 
    }
    \label{fig:FXCDM_BAO}
\end{figure}

We present our key findings in Figures~\ref{fig:Fisher_Not_OK_1}-\ref{fig:FXCDM_BAO}. For both the flat $\Lambda$CDM and flat $w$CDM models, the Fisher approximation and DALI approximation to the likelihood level curves show significant differences.
This can be seen very clearly in Figure~\ref{fig:Fisher_Not_OK_1}, which shows the results of the cross-section test applied to the flat $\Lambda$CDM model with QSO data. In the top right panel, the 1- and 2-$\sigma$ confidence contours derived from the Fisher approximation are shown as purple dotted curves, the 1- and 2-$\sigma$ confidence contours derived from the DALI approximation are shown as green dashed curves, and the red dots represent points in the parameter space at which contours of constant likelihood are concave, $K < 0$. As discussed in Sec. \ref{sec:Curvature}, if the confidence contours in a region of parameter space have negative extrinsic curvature, then the confidence contours are concave there. From the marginalized contours shown in the triangle plot, we can see that the exact likelihood and the DALI approximation are concave, and that the concavity points in the direction of increasing $H_0$, and slightly in the direction of increasing $\Omega_{m}$, matching what we see in the upper panel of Figure \ref{fig:Fisher_Not_OK_1}. A similar conclusion can be drawn from Figure \ref{fig:Fisher_Not_OK_2}, which shows the results of the cross-section test and MCMC sampling applied to the flat $\Lambda$CDM model with $H(z)$ data. In this case, because the $H(z)$ data give tighter constraints on the cosmological model parameters, the difference between the exact and DALI contours is less pronounced. Nevertheless, the cross-section test indicates that the Fisher approximation is not fully valid in parts of the parameter space, and comparison with the exact likelihood confirms the breakdown of the Fisher approximation.
In contrast, Figure \ref{fig:Fisher_OK} shows a case in which the Fisher approximation works well, at least out to 2$\sigma$. In this figure, the upper right panel displays the results of the cross-section test applied to the flat $\Lambda$CDM model with BAO data, and the triangle plot displays the marginalized confidence contours obtained from MCMC sampling of the same model and data. From the triangle plot, we can see no difference between the exact likelihood or the DALI approximation, and the confidence contours of both align closely with the confidence contours of the Fisher approximation. From the upper panel of Figure~\ref{fig:Fisher_OK}, we can see that while there is negative extrinsic curvature in some regions of the parameter space, these regions do not overlap with the 2$\sigma$ confidence contours, and the Fisher approximation and DALI approximation predict similar level curves of the likelihood. The Fisher approximation is therefore adequate within 2$\sigma$ of the best-fit point.

Figure~\ref{fig:FXCDM_Hz} shows an application of the cross-section test to the 3-parameter flat $w$CDM model, along with a triangle plot showing the marginalized contours obtained from MCMC sampling. The panels in the upper right triangle of this figure show 2-dimensional cross-sections of the Fisher and DALI approximations to the flat $w$CDM likelihood. In each of these upper right panels, the model parameter not shown is held fixed at its best-fit value obtained from the fit to the exact likelihood. For example, in the top middle panel, $w = -1.3$ while $\Omega_{m}$ and $H_0$ are left free to vary. In these cross-section panels, as in Figures~\ref{fig:Fisher_Not_OK_1}-\ref{fig:Fisher_OK}, purple dotted curves represent the likelihood contours derived from the Fisher approximation, and green dashed curves represent the likelihood contours derived from the DALI approximation. Red dots indicate regions of the parameter space in which $K < 0$. The main difference between the flat $w$CDM and the flat $\Lambda$CDM results is that, because we apply the cross-section test to slices of the (approximate) likelihood for flat $w$CDM, it is not as easy to compare these results to what we would expect from an inspection of the marginalized constraint contours shown in the lower left panels of Figure~\ref{fig:FXCDM_Hz}. Nevertheless, we can see from the lower left panels that the Fisher approximation provides a poor approximation of the exact likelihood, and while the DALI approximation is not significantly better, it does roughly capture the shape of the exact likelihood. The concavity of the DALI contours is evident in the $\Omega_m$ - $H_0$ and $\Omega_m$ - $w$ panels of Figure \ref{fig:FXCDM_Hz}, and the output of the curvature calculation reflects this.

Figure~\ref{fig:FXCDM_BAO} shows the marginalized and cross-sectional constraint contours for a flat $w$CDM model, though here the model is fitted to BAO data only.  The unmarginalized cross-sections from the DALI and Fisher approximations agree well to 2$\sigma$, except for the $H_0$ - $w$ cross-section, which shows noticeable disagreement between DALI and Fisher even at 1$\sigma$.  
This example demonstrates that one should be wary of the Fisher approximation when the DALI approximation to the likelihood differs from the Fisher approximation in the vicinity of the best-fit point, even if only for a subset of parameters.  

\subsection{CMB forecasts}
\label{subsec:high-z_results}

\begin{figure*}[htbp]
    \centering
    \includegraphics[width=2.0\columnwidth]{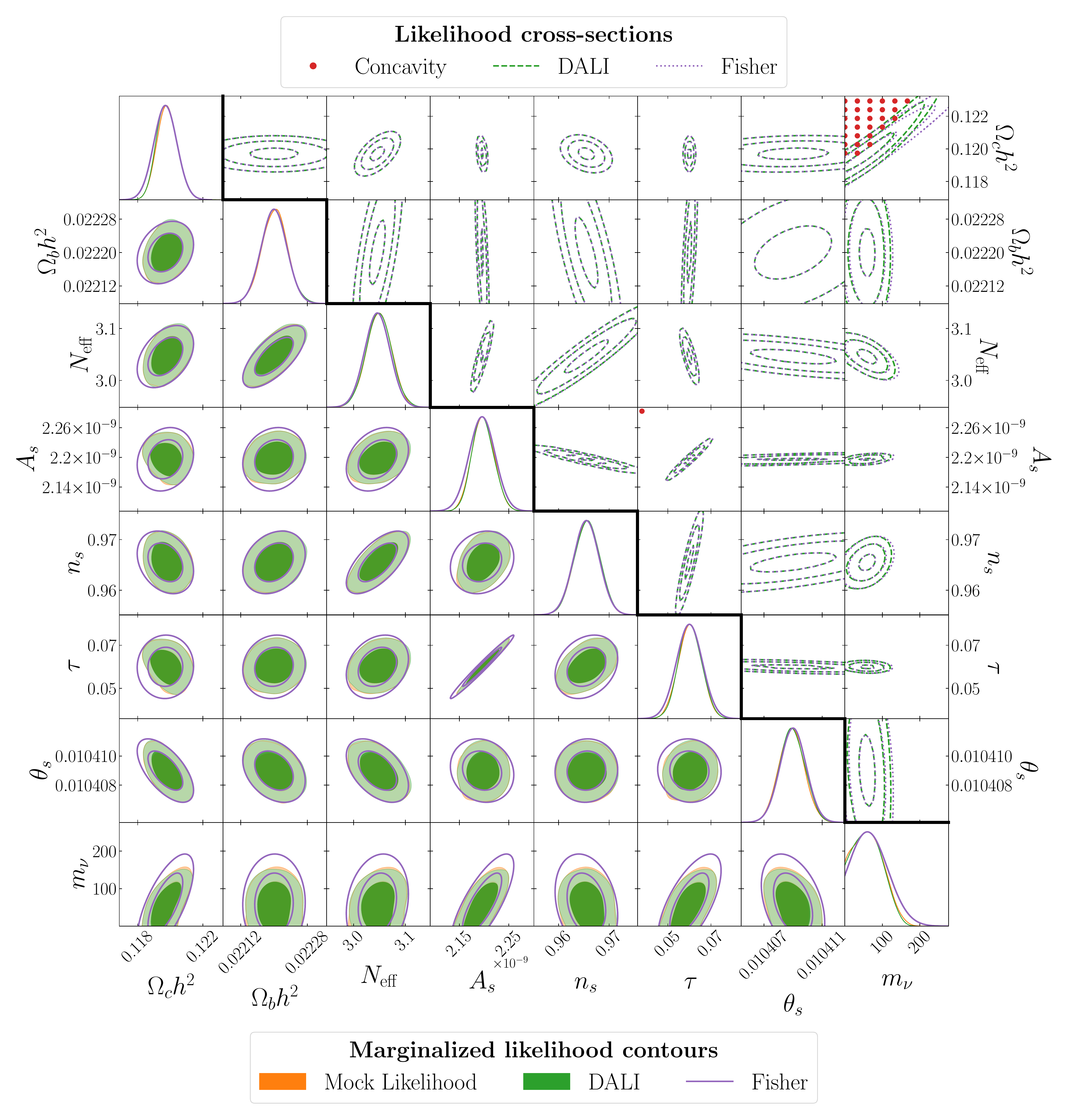}
    \caption{Constraint contours for CMB Experiment B, from the three methods discussed in the text. The fiducial value of $m_\nu$ is taken to be 60 meV, and BAO data is not included. In the top-right panel, one can see the negative extrinsic curvature present in the $\Omega_c h^2$ - $m_{\nu}$ cross-section. This is consistent with the lower-left panel, where one can see that the constraint contours from the DALI and mock likelihood predictions are curved, and they do not agree well with the Fisher forecast. Although they do not show any points of negative extrinsic curvature in the likelihood cross-sections, the other $m_{\nu}$ marginalized contours also show disagreement between the results from the Fisher method, and the DALI and mock likelihood methods, for the reasons discussed in the text. Neutrino masses are reported in meV.
    }
    \label{fig:ExpB_CMB_60meV}
\end{figure*}

Figure~\ref{fig:ExpB_CMB_60meV} shows results for CMB forecasts for Experiment B defined in Table~\ref{table:experiments}, using the methods described in Section~\ref{subsec:CMB_forecasts}.
The cross-section with the most pronounced deviation from an ellipsoidal shape is the $\Omega_c h^2$ - $m_{\nu}$ cross-section, shown in the top-right panel. This is consistent with a general pattern we can see by examining Figure~\ref{fig:ExpB_CMB_60meV}.  Specifically, we find that there is noticeable disagreement between the Fisher forecast and the mock likelihood forecast for the two-dimensional marginalized parameter constraints involving $m_\nu$ and $\Omega_c h^2$. We find one isolated point of negative extrinsic curvature in the $A_s$ - $\tau$ cross-section, but it is located far from the 2$\sigma$ region of interest, and the Fisher and DALI cross-sections match closely. Accordingly, we find that the Fisher-approximated marginalized contour for $A_s$ and $\tau$ matches well with that obtained from the mock likelihood.

There is one more source of discrepancy between the Fisher and mock likelihood forecasts that results not from the breakdown of the Fisher approximation, but rather from the marginalization procedure.  Marginalized Fisher constraints are computed by simply inverting the Fisher matrix, which implicitly treats all parameters as if they take values from $-\infty$ to $+\infty$.  However, sampling the mock likelihood is only possible in regions where parameters take physically meaningful values.  This makes little difference for parameters whose fiducial values are many standard deviations away from the non-physical regime.  However, in the case considered here, $m_\nu$ is physically restricted to be non-negative, and the fiducial value of $60$~meV lies only about 1$\sigma$ from 0.  As a result, marginalized constraints obtained with the mock likelihood (and the DALI approximation) show deviation from the Fisher forecast for parameters that are degenerate with $m_\nu$.  This causes some non-ellipticity of the marginalized contours involving $A_s$ and $\tau$ on the low end of each of these parameters.  One could choose to also marginalize the Fisher forecasts by sampling the Fisher-approximated likelihood, restricting to the same parameter range.  We tested this procedure and found in that case the difference in the low end of the marginalized constraints on $A_s$ and $\tau$ between the Fisher forecast and the mock likelihood disappeared, though the disagreement on the high end of these parameters (and on $m_\nu$) resulting from concavity of the likelihood contours remained.

\begin{table*}[htbp!]
\renewcommand{\arraystretch}{1.2}
\begin{center}
 \begin{tabular}{l @{\hskip 12pt} c @{\hskip 12pt} |@{\hskip 12pt} c @{\hskip 12pt} |@{\hskip 12pt} c @{\hskip 12pt} |@{\hskip 12pt}c@{\hskip 12pt} } 
 \toprule
   Configuration & Fiducial $m_\nu$    &   Fisher & DALI & Mock Likelihood \\ [0.5ex] 
 \hline
Exp A         & 60  & $60\pm60$ & $66^{+27}_{-55}$ & $70^{+30}_{-57}$ 	\\

Exp A w/BAO   & 60  & $60\pm27$ & $59\pm 25$ & $60\pm 27$	\\

Exp A      &  200   & $200\pm41$ & $200^{+43}_{-38}$ & $200^{+45}_{-39}$	 \\ 

Exp B        &  60  & $60\pm54$ & $64^{+30}_{-49}$ & $65^{+29}_{-53}$   \\

Exp B w/BAO  & 60  & $60\pm25$ & $59\pm 24$ & $60^{+28}_{-24}$        \\

Exp B        & 200  & $200\pm32$ & $200^{+34}_{-30}$ & $200^{+34}_{-29}$   \\
  \hline
\end{tabular}
    \caption{
    Forecasted upper and lower 68\% confidence intervals for $m_\nu$ in meV derived using the Fisher matrix, the DALI technique, and a mock likelihood. Fisher gives a larger interval than DALI or the mock likelihood when the fiducial value of $m_\nu$ is $60$~meV, as discussed in more detail in the text. The three forecasting techniques yield similar confidence intervals when one sets the fiducial value of $m_\nu$ to be more than 1$\sigma$ away from $0$~meV, or when BAO data is included to further tighten forecast constraints.
    }
\label{table:lcdm_sigmas_mnu}
\end{center}
\end{table*}

In Table~\ref{table:lcdm_sigmas_mnu}, we present the upper and lower 68\% confidence intervals for $m_\nu$ for several sets of CMB forecasts. 
One might be surprised by the fourth line, corresponding to the results presented in Figure~\ref{fig:ExpB_CMB_60meV}, which shows that the Fisher forecast produces larger errors than the DALI and mock likelihood forecasts.
This behavior can be understood in terms of the non-ellipticity discussed above. 
(Similar behavior can be seen in Figure~\ref{fig:FXCDM_Hz}.)
In Table~\ref{table:lcdm_sigmas_mnu} we show two additional forecasts involving Experiment B, where the Fisher forecasted errors closely match those from the DALI and mock likelihood treatments.
Firstly, we show the effect of including BAO constraints in the forecast, which tightens $m_\nu$ contours and results in similar constraints predicted by all three forecasting methods. One can also change the fiducial value of $m_\nu$, for example to 200 meV, which shifts the constraints away from the $m_\nu = 0$ cutoff and the contours regain their elliptical, Gaussian shapes, with constraints once again being similar across each of the three methods.

\subsection{Primordial abundance measurements}
\label{subsec:BBN_model}

\begin{figure}
    \centering
    \includegraphics[width=\columnwidth]{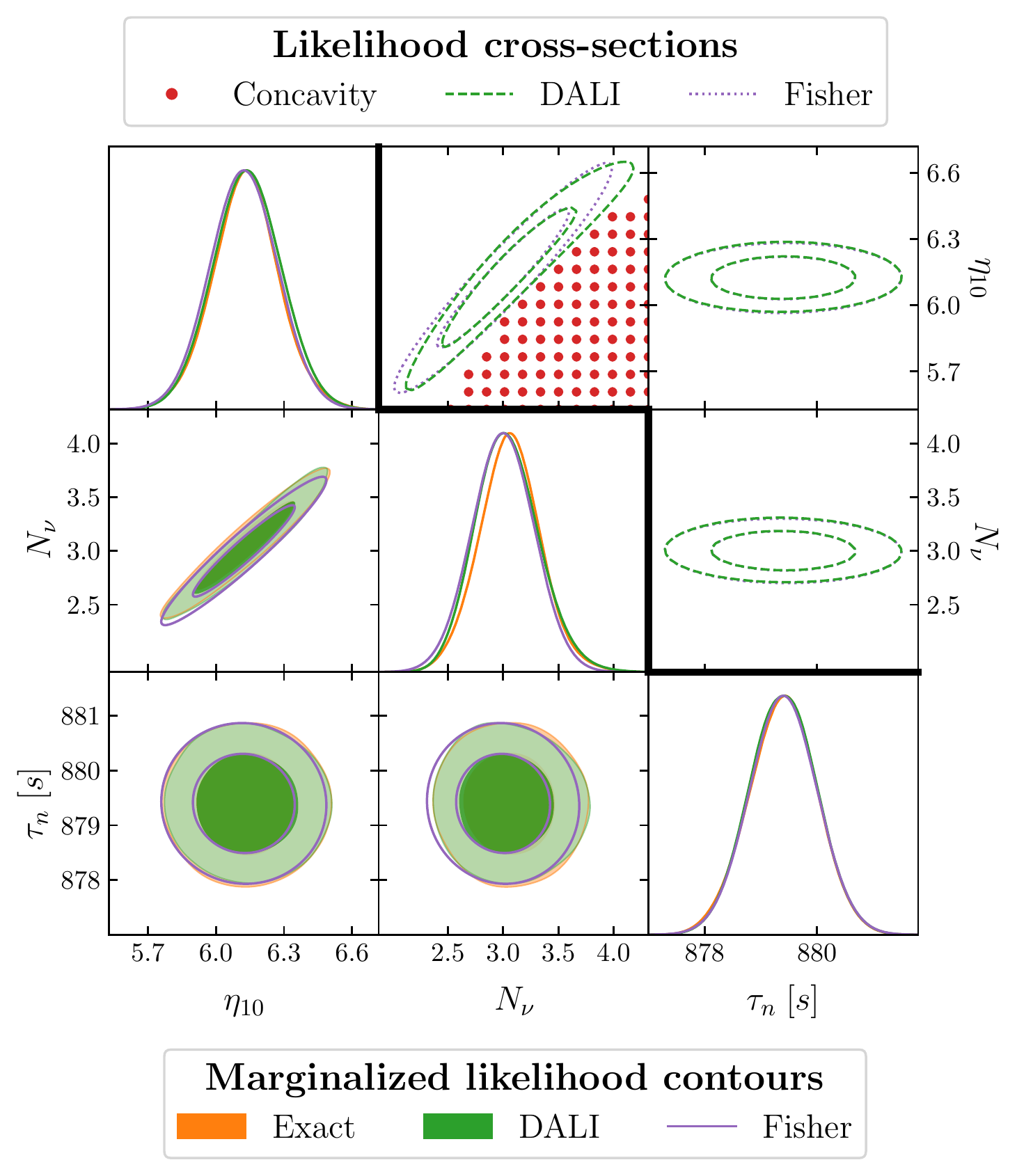}
    \caption{
    Parameter constraints from primordial abundance measurements assuming $\sigma(\tau_n) = 0.6$~s in accordance with the current global average~\cite{ParticleDataGroup:2020ssz}. Concavity of the DALI contour in the top middle panel suggests that the Fisher approximation may be suspect over some range of parameter space.  The marginalized 2-$\sigma$ contours of the exact likelihood shown in the bottom panels show a very slight deviation from the Fisher approximation.
    }
    \label{fig:BBN_06}
\end{figure}

\begin{figure}
    \centering
    \includegraphics[width=\columnwidth]{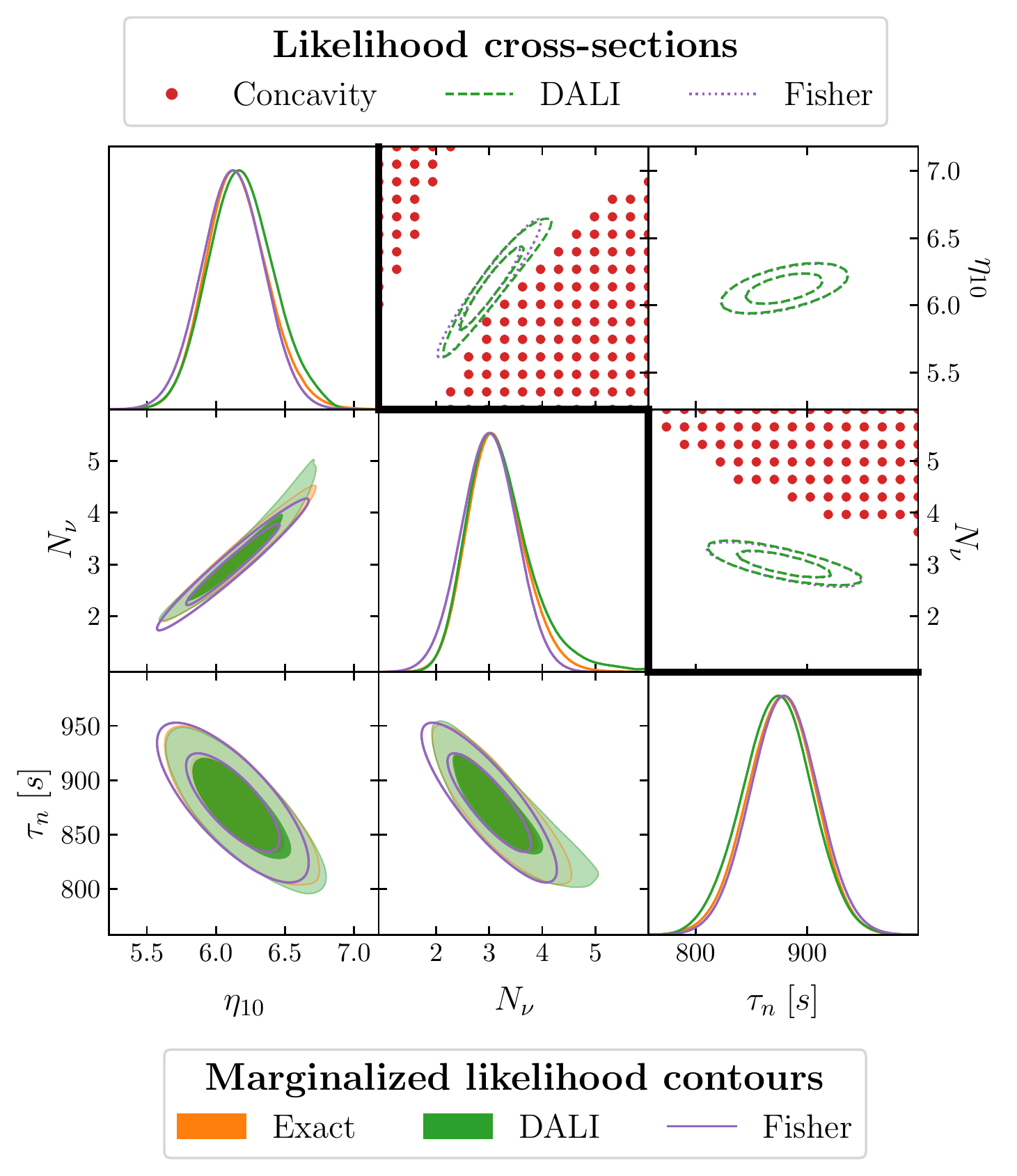}
    \caption{Same as Figure~\ref{fig:BBN_06} but with $\sigma(\tau_n) = 30$~s. One can see that as observational constraints are weakened, the deviation from the Fisher approximation becomes more prominent.
    }
    \label{fig:BBN_30}
\end{figure}

The results of the analysis with primordial abundance measurements, detailed in Section~\ref{subsec:BBN_forecasts}, are shown in Figures~\ref{fig:BBN_06} and \ref{fig:BBN_30}. 
The plots use the same observational uncertainties on $\isotope{D}/\isotope{H}$ and $Y_p$, but different uncertainties on the neutron lifetime $\tau_n$.

Figure~\ref{fig:BBN_06} shows the results with  $\sigma(\tau_n)=0.6$~s, which is the current global average~\cite{ParticleDataGroup:2020ssz}. 
The concavity of the cross-section of the DALI approximation to the likelihood shown in the $N_\nu$ - $\eta_{10}$ panel shows that the Fisher approximation is near the edge of its applicability, especially as one moves toward the tails of the distribution.  This is reflected in the marginalized likelihood contours, which show that the 1-$\sigma$ constraints agree well among the Fisher, DALI, and exact likelihood treatments, though there are slight differences among the 2-$\sigma$ contours.
Figure~\ref{fig:BBN_30} shows the results of the same analysis but with a much weaker constraint on the neutron lifetime of $\sigma(\tau_n) = 30$~s.  All parameters are more weakly constrained in this case, and the deviation of the Fisher approximation from the exact likelihood becomes more prominent. 

These cases show concavity in the DALI-approximated likelihood cross-sections, suggesting that the Fisher approximation may not be valid in some parts of parameter space. However, one can see that the deviation from ellipticity is small, and the level curves of the Fisher and DALI approximations do not significantly differ.  One would therefore expect that the Fisher approximation provides a good approximation to the exact likelihood, which we can see is borne out in the comparison of the marginalized constraint contours at least out to $2\sigma$.


\section{Limitations of DALI Approximation}
\label{sec:Limitations_of_DALI}

We have demonstrated how two-dimensional slices of the DALI approximation to the likelihood level curves can be used to detect where the Fisher approximation breaks down.  One might reasonably wonder whether Fisher forecasts should simply be replaced with the DALI approximation for all parameter forecasts, given that our cross-section test requires the computation of the relevant ingredients for a DALI forecast.  In this section we discuss some reasons to be cautious about inferring forecasted parameter constraints from the DALI formalism, at least without careful validation of the results.

\subsection{Numerical / convergence issues}

\begin{table}[htbp]
\renewcommand{\arraystretch}{1.2}
    \centering
    \begin{tabular}{l @{\hskip 12pt}  c @{\hskip 12pt} c @{\hskip 12pt} c}
    \hline
    \hline
        Run & Parameter & Exact &  DALI  \\
        \hline
         \multirow{2}{*}{1} &  $H_0$ & 1.08 &  1.49\\
         & $\Omega_{m}$ & 0.000287 & 0.000879\\
         \hline
         \multirow{2}{*}{2} & $H_0$ & 1.08 &  1.10\\
         & $\Omega_{m}$ &  0.000287 & 0.000283\\
          \hline
         \multirow{2}{*}{3} & $H_0$ & 1.10 &  1.23\\
         & $\Omega_{m}$ &  0.000288 & 0.000548\\
          \hline
         \multirow{2}{*}{4} & $H_0$ & 1.11 &  1.10\\
         & $\Omega_{m}$ &  0.000289 & 0.000414\\
         \hline
         \hline
         \multirow{2}{*}{Fisher} &   $H_0$  & \multicolumn{2}{c}{1.00} \\
         & $\Omega_{m}$ & \multicolumn{2}{c}{0.000270}  \\
         \hline
    \end{tabular}
    \caption{Variances of parameters of flat $\Lambda$CDM model fitted to BAO data, computed with the exact likelihood,  the DALI approximation of the likelihood, and the Fisher approximation of the likelihood. 
    We used a convergence criterion of $R-1 = 5\times 10^{-6}$. Across all four runs, the variances computed with the exact likelihood are fairly stable, while the variances computed with the DALI approximation exhibit larger-amplitude fluctuations.
    }
    \label{tab:FLCDM_BAO_var_extended3}
\end{table}

Obtaining marginalized constraints in the DALI formalism requires numerical integration of the approximated likelihood.  Compared to evaluation of the exact likelihood, the DALI formalism can offer a significant reduction in computation time, since it requires many fewer evaluations of the likelihood.  However, this numerical integration, even if it is faster, introduces some undesirable sensitivity to numerical details.  For example, MCMC techniques often rely upon a convergence criterion related to the Gelman-Rubin statistic $R-1$ comparing the variance of means among Markov chains to the mean of the variance of samples within each chain~\cite{Gelman:1992abc}.  The choice of this convergence criterion can often impact the results that will be found in a given case.  One can typically produce more stable and repeatable results at the cost of extra computational time by requiring a tighter convergence criterion, though this can become quite costly when one requires robust predictions more than a few $\sigma$ away from the best-fit point.

One might hope that summary statistics, like the variance of a given marginalized parameter constraint, might be mostly insensitive to numerical details.  However, we find that this is not reliably the case in our DALI forecasts.  From Figure \ref{fig:Fisher_OK}, we can see that for the flat $\Lambda$CDM model with BAO data, both the DALI and the Fisher approximations appear to agree with the constraint contours of the exact likelihood (at least to within 2$\sigma$). If we calculate the variances of the model parameters using the exact, Fisher, and DALI likelihoods, however, we find less agreement than what seems to be implied by the figure. Table~\ref{tab:FLCDM_BAO_var_extended3} shows the results of these calculations. We computed the variances of the model parameters, using both the exact likelihood and the DALI approximation, in several runs. We found that while the variances of the parameters computed using the exact likelihood exhibited relatively little variation, the variances computed using the DALI approximation can show as much as a 61\% deviation from the variance averaged across all runs (in the case of $\Omega_{m}$ in run 1).

Another sensitivity to numerical details arises in the treatment of priors.  Numerical integration requires that some choices be made about the range over which to sample parameters.  One will generally find different marginalized constraints for different choices of priors, and it is not always obvious which choice should be made.  For example, when model parameters are subject to physical constraints (such as for parameters that are only physically meaningful when greater than or equal to zero), it might seem natural to place this restriction on the numerical integration.  However, this treatment of priors will lead to marginalized constraints that should not be compared directly to what is obtained by inverting the Fisher matrix (which formally integrates all parameters over all real numbers).  This discussion does not imply that anything is wrong with any particular choice of priors used when carrying out numerical integration, but the fact that the results will depend on that choice makes the resulting forecasts less transparent.

\subsection{Toy Models}

Though it may be possible to overcome the numerical issues discussed in the previous subsection, we now turn to cases in which the DALI approximation to the likelihood is worse than the Fisher approximation.
We focus on two toy models, both of which depend on a single parameter $a$. In the first model, the observable $x$ takes the form $x = a^2$, and in the second model, $x$ takes the form $x = \left(1 + a^2\right)^{1/4}$. The key difference between these two models is that the first exhibits super-linear (quadratic) dependence on $a$, while the second exhibits sub-linear dependence on $a$.\footnote{For a discussion of sub-linear parameter dependence in the context of Fisher forecasts see, e.g.~Ref.~\cite{Lee:2021bmn}.}

\begin{figure}[t!]
    \centering
    \includegraphics[width=1.0\columnwidth]{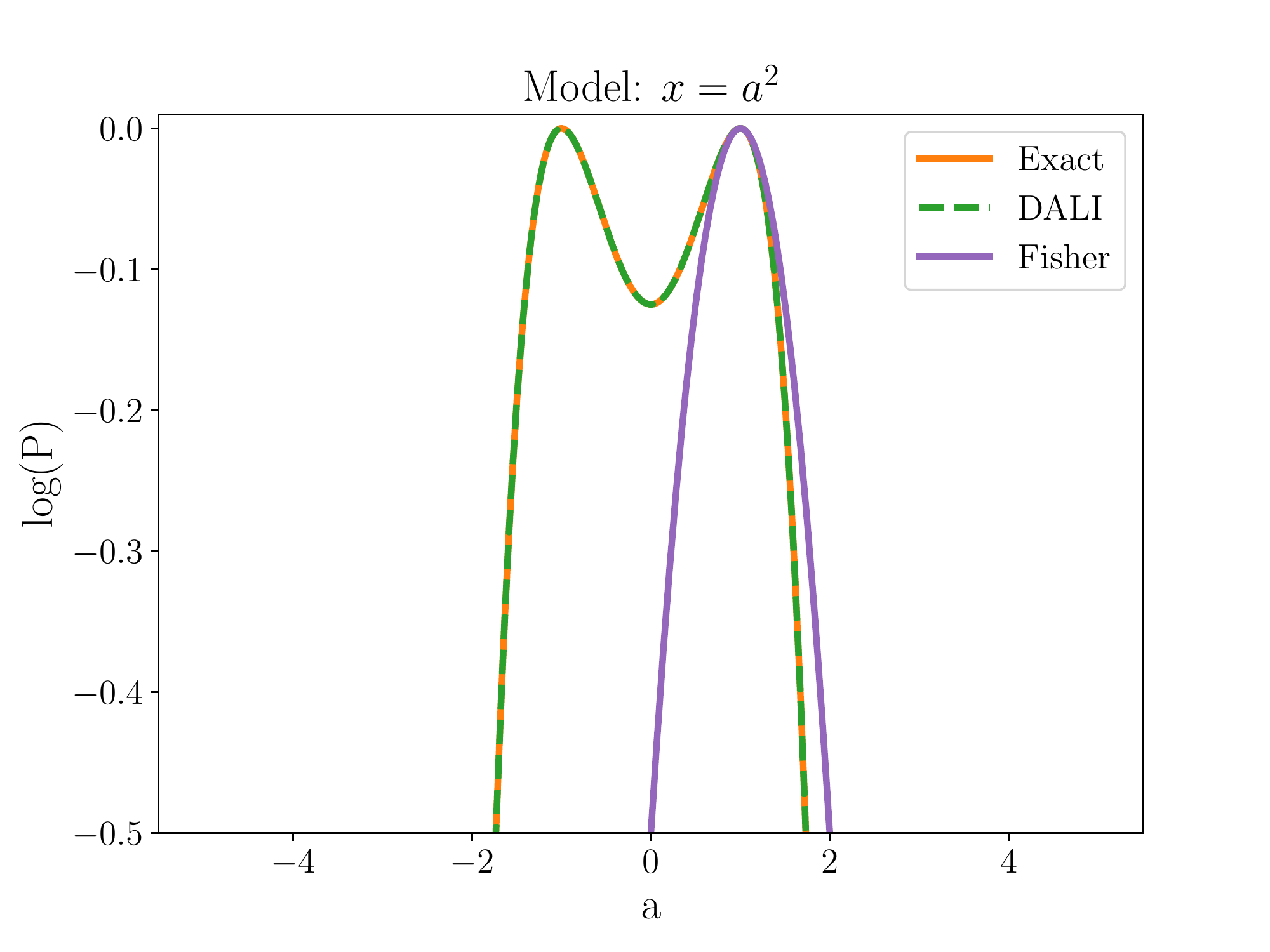}
    \caption{Toy model of the form $x=a^2$. The assumed uncertainty on $x$ is $\sigma_x = 2$, and the fiducial value of $a$ is taken to be $a = 1$.}
    \label{fig:ToyModel_a_2_1}
\end{figure}

\begin{figure}[t!]
    \centering
    \includegraphics[width=1.0\columnwidth]{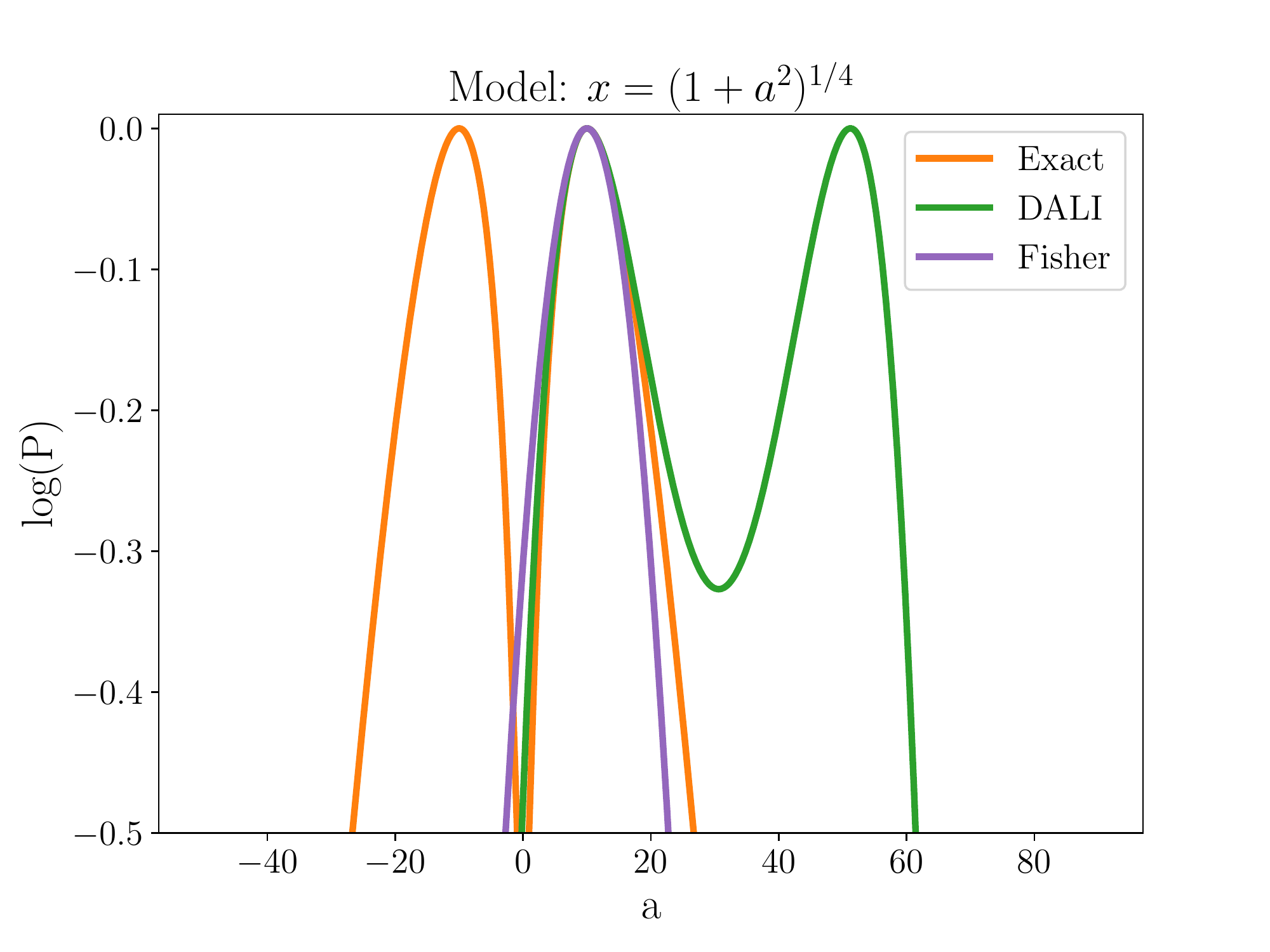}
    \caption{Toy model of the form $x=(1+a^2)^{1/4}$. The assumed uncertainty on $x$ is $\sigma_x = 2$, and the fiducial value of $a$ is taken to be $a = 10$.}
    \label{fig:ToyModel_a_2_4}
\end{figure}

We tested the behavior of both the Fisher and DALI approximations against that of the exact likelihood for each model by plotting ${\rm log}(P)$ versus $a$, where $P$ is the posterior. These plots are shown in Figures \ref{fig:ToyModel_a_2_1} and \ref{fig:ToyModel_a_2_4}. From Figure~\ref{fig:ToyModel_a_2_1}, we can see that the DALI approximation matches the exact likelihood perfectly when the given model has quadratic parameter dependence. The Fisher approximation, as expected, only works well in the vicinity of one of the critical points of ${\rm log}(P)$. Figure~\ref{fig:ToyModel_a_2_4}, by contrast, shows a very different picture. For our sub-linear model, both the Fisher and DALI approximations describe the exact likelihood well near one of the critical points of ${\rm log}(P)$, but they both break down far from the critical point.  Furthermore, the DALI approximation to the likelihood exhibits a spurious peak at large values of $a$ that does not appear in the Fisher approximation.  We can see that the DALI approximation to the likelihood is strictly worse than the Fisher approximation for that region of parameter space in this model.

\section{Conclusion}
\label{sec:Conclusion}

We have developed a scheme to test the validity of the Fisher approximation that is based on the DALI formalism.  We showed how two-dimensional slices of the level curves of the DALI approximation to the likelihood can be used to detect the breakdown of the Fisher approximation.


We applied our cross-section test to three cosmological data sets: expansion history data, forecasted CMB observations, and primordial light element abundance measurements. Cases that exhibited concavity of likelihood contours and significant differences between the cross sections of the DALI-approximated and Fisher-approximated likelihood level curves typically showed significant deviation of the Fisher approximation from the true likelihood, suggesting that Fisher forecasts were insufficient in these cases.  While we focused here on cosmological forecasting, the techniques we developed are applicable to any statistical forecast where Fisher forecasting can be employed.

In the cases where the Fisher formalism breaks down, we found that the DALI forecasts sometimes provide a better match to the true likelihood.  However, we do not generally recommend that DALI forecasts should replace Fisher forecasts.  We discussed how numerical issues can complicate interpretation of DALI forecasts and also how DALI forecasts in some cases perform worse than the corresponding Fisher forecasts.  We therefore recommend that the cross-section test described here be used as a means to check the validity of Fisher forecasts, and that one should instead rely on sampling of the true likelihood in cases where the Fisher formalism breaks down.

Fisher forecasts are by no means perfect and are built on assumptions that we often know to be violated at some level by real observations.  However, the efficiency with which the Fisher matrix can be computed and the simplicity of its interpretation ensure that the Fisher formalism will continue to hold a prominent role in forecasting.  The test we developed here provides a means to test the validity of Fisher forecasts in any given scenario, and its application will help to discriminate the cases where we should be reasonably confident in Fisher forecasts from those in which a more complete treatment is required.

\section*{Acknowledgments}
The authors would like to thank Nicola Bellomo, Connor Hayward, Bohua Li, and Paul Shapiro for helpful discussions.
This work is supported in part by the US~Department of Energy under Grant~\mbox{DE-SC0010129}.
Computational resources for this research were provided by SMU’s Center for Research Computing.

\bibliographystyle{utphys}
\bibliography{DALI}

\end{document}